\documentclass[prodmode,acmcsur]{acmsmall} 

\usepackage[ruled]{algorithm2e}
\usepackage{rotating}
\usepackage{booktabs}

\renewcommand{\algorithmcfname}{ALGORITHM}
\SetAlFnt{\small}
\SetAlCapFnt{\small}
\SetAlCapNameFnt{\small}
\SetAlCapHSkip{0pt}
\IncMargin{-\parindent}

\usepackage{todonotes}
\usepackage{soul}
\newcounter{todocounter}
\begin{document}

\markboth{Santana, E. F. Z. et al.}{Software Platforms for Smart Cities}

\title{Software Platforms for Smart Cities: \\Concepts, Requirements, Challenges, and a Unified Reference Architecture}
\author{Eduardo Felipe Zambom Santana
\affil{University of S\~ao Paulo}
Ana Paula Chaves
\affil{Federal Technological University of Paran\'a}
Marco Aurelio Gerosa
\affil{University of S\~ao Paulo}
Fabio Kon
\affil{University of S\~ao Paulo}
Dejan S. Milojicic
\affil{Hewlett Packard Labs Palo Alto}
}

\begin{abstract}
Making cities smarter help improve city services and increase citizens' quality of life. Information and communication technologies (ICT) are fundamental for progressing towards smarter city environments. Smart City software platforms potentially support the development and integration of Smart City applications. However, the ICT community must overcome current significant technological and scientific challenges before these platforms can be widely used. This paper surveys the state-of-the-art in software platforms for Smart Cities. We analyzed 23 projects with respect to the most used enabling technologies, as well as functional and non-functional requirements, classifying them into four categories: Cyber-Physical Systems, Internet of Things, Big Data, and Cloud Computing. Based on these results, we derived a reference architecture to guide the development of next-generation software platforms for Smart Cities. Finally, we enumerated the most frequently cited open research challenges, and discussed future opportunities. This survey gives important references for helping application developers, city managers, system operators,
end-users, and Smart City researchers to make project, investment, and research decisions.
\end{abstract}


\terms{Digital Cities, Internet of Things, Big Data, Cloud Computing, Cyber-Physical Systems, Middleware, Infrastructure}

\keywords{Wireless sensor networks, Software platforms, Middleware, Infrastructure}

\maketitle

\section{Introduction}
\label{sec:introduction}

Since 2009, most of the world's population lives in cities \cite{un2009urbanRural}. Current resources and infrastructure are hardly enough to cope with the increasing demand that population growth and geographic concentration generates \cite{caragliu2011smart}. Making cities smarter can help optimize resource and infrastructure utilization toward increased sustainability. One approach involves creatively combining the large amounts of data generated by multiple city sources (such as sensor networks, traffic systems, user devices, and social networks) to create integrated services and applications, improving city services, and making better use of city resources. However, efficiently and effectively using all these data sources is a challenge.

Initiatives for developing Smart City systems have been proposed in a wide range of city services, such as transportation \cite{smartCityTransportation}, traffic control \cite{barba2012smart}, air pollution \cite{vakali2014datastream}, waste management \cite{perera2014iot}, health care \cite{HussainHealthCare}, public safety \cite{cloutPlatform}, water \cite{perez2015public}, energy \cite{yamamoto2014using}, and emergency management \cite{smartCityDisasterManagement}. However, most of these solutions focus on a specific domain, target a specific problem, and were developed from scratch, with little software reuse. They do not interoperate, leading to duplication of work, incompatible solutions, and non-optimized resource use.



Integrating all of these domains into a complete and consistent solution require basic services from the underlying software infrastructure. Such basic services could be provided by a novel, comprehensive software platform, which could include facilities for application development, integration, deployment, and management, easing the construction of sophisticated Smart Cities applications. We define a software platform for Smart Cities as

\begin{center}
\textit{``an integrated middleware environment that supports software developers in designing, implementing, deploying, and managing applications for Smart Cities.''}
\end{center}

Many challenging issues still need to be addressed before a highly effective software platform for Smart Cities can be created, including: enabling interoperability between a city's multiple systems, guaranteeing citizens' privacy, managing large amounts of data, supporting the required scalability, and dealing with a large variety of sensors.

In the research described in this paper, we evaluated initiatives for developing software platforms for Smart Cities, aiming to comprehensively analyze relevant functional and non-functional requirements, according to the literature. 
Based on the analysis, we derived a reference architecture that addresses these requirements. With this survey, we intend to clarify important aspects of the design, development, and management of Smart Cities platforms. To do so, we examined 23 Smart Cities software platforms, aiming to answer the following \textbf{general research question}:

\begin{center}
\textit{What is required for the development of a 
software platform for enabling the construction of 
scalable integrated Smart City applications?}
\end{center}

We investigated three more specific research questions:

\begin{description}
\item[RQ1:] ``What are the enabling technologies used in state-of-the-art software platforms for Smart Cities?''
\item[RQ2:] ``What are the requirements that a software platform for Smart Cities should meet?''
\item[RQ3:] ``What are the main challenges and open research problems in the development of next generation robust software platforms for Smart Cities?''
\end{description}

To answer research question RQ1, we identified the most common enabling technologies employed in platforms for Smart Cities. As described in Section \ref{subsec:tech-building-blocks}, we grouped them into four main categories: \textbf{Internet of Things (IoT)} \cite{atzori2010iotsurvey}, applied to control sensors and actuators responsible for retrieving information from the city; \textbf{Big Data} \cite{mayer2013big}, to support storage and processing of the data collected from the city; \textbf{Cloud Computing} \cite{armbrust2010view}, to provide elasticity to the services and data storage; and \textbf{Cyber-Physical Systems} \cite{white2010cps}, to enable the interaction of systems with the city environment. To answer RQ2, we identified the most common functional and non-functional requirements for developing a platform for Smart Cities, as described in Section \ref{subsec:requirements}. Finally, to answer RQ3, we explored the main challenges researchers identified in developing software platforms for Smart Cities, as discussed in Section \ref{sec:challenges}.

Combining the results of the three research questions, we derived a reference architecture. This architecture presents components for implementing a software platform for Smart Cities, based on the most common enabling technologies, the requirements, and challenges surveyed in this research. We also discuss the critical implications of platforms for Smart Cities in the Section \ref{sec:discussion}.

The remainder of this paper is organized as follows. Section \ref{sec:main-concepts} presents the definition of Smart Cities and introduces the four enabling technologies for platforms for Smart Cities. Section \ref{sec:software-platforms} presents the platforms, architectures, and implemented systems for Smart Cities, grouped according to the enabling technologies that each platform uses. Section \ref{sec:challenges} points out challenges and open research problems in the development of a platform for Smart Cities. In Section \ref{sec:reference-architecture}, we present a reference architecture for software platforms for Smart Cities. In Section \ref{sec:discussion},  we discuss the relationship between the requirements and the enabling technologies as well as their implications for the development of software platforms for Smart Cities. Section \ref{sec:related-works} presents the related work and, finally, Section \ref{sec:conclusions} presents our conclusions.

\section{Main Concepts}
\label{sec:main-concepts}

We now introduce the main concepts used in the discussions within this survey. We first present definitions of Smart Cities and, then, discuss the most adopted enabling technologies for the development of software platforms for Smart Cities.

\subsection{Smart Cities}
\label{subsec:smart-cities-definitions}

The term ``Smart City'' has many different definitions. Some exceed the software context, focusing only on social or business aspects. Regarding software systems, many authors define a Smart City as the integration of social, physical, and IT infrastructure to improve the quality of city services \cite{caragliu2011smart,hollands2008will}. Other authors focus on a set of Information and Communication Technology (ICT) tools used to create an integrated Smart City environment \cite{hollands2008will,washburn2009helping,bowerman2000vision}.

Giffinger et al. \cite{giffinger2007smart} assert that a Smart City has six main dimensions: smart economy, smart people, smart governance, smart mobility, smart environment, and smart living. Many authors adopt this definition \cite{munoz2011forefront,papa2013towards} and there are even benchmarks to produce a ranking of the smartest city using these dimensions\footnote{Smarts Cities in Europe - \url{http://www.smart-cities.eu}}.

In their definition of Smart Cities, Washburn et al. \cite{washburn2009helping} and Hall et al. \cite{bowerman2000vision} emphasize integrating software services and applications to improve regular city services and the lives of their citizens. Following this idea, Kanter and Litow \cite{litow2009scmanifesto} declare that creating independent software for each city domain is not sufficient for creating an environment for Smart Cities. They contend that all city sub-systems (such as transport, education, energy, and water) must be linked in a network as an organic whole to provide integration among all city subsystems. Caragliu et al. \cite{caragliu2011smart} definition of Smart Cities highlights the significant benefit of sustainability and management of natural resources.

We are aligned with the vision that a city must have an integrated environment to facilitate the interoperability between the city's sub-systems. Based on that, in our view:

\begin{center}
\textit{``a Smart City is a city in which its social, business, and technological aspects are supported by Information and Communication Technologies to improve the experience of the citizen within the city. To achieve that, the city provides public and private services that operate in an integrated, affordable, and sustainable way.''}
\end{center}

To make a city smarter, it is desirable to integrate services and applications in a unified technological infrastructure. A sensible way to make the above reality is with a well-designed software platform providing the necessary infrastructure for dealing with large volumes of data, a wide variety of devices and applications, system interoperability, and other problems related to Smart City environments.

\begin{figure}[!htb]
\centering
\includegraphics[height=6cm]{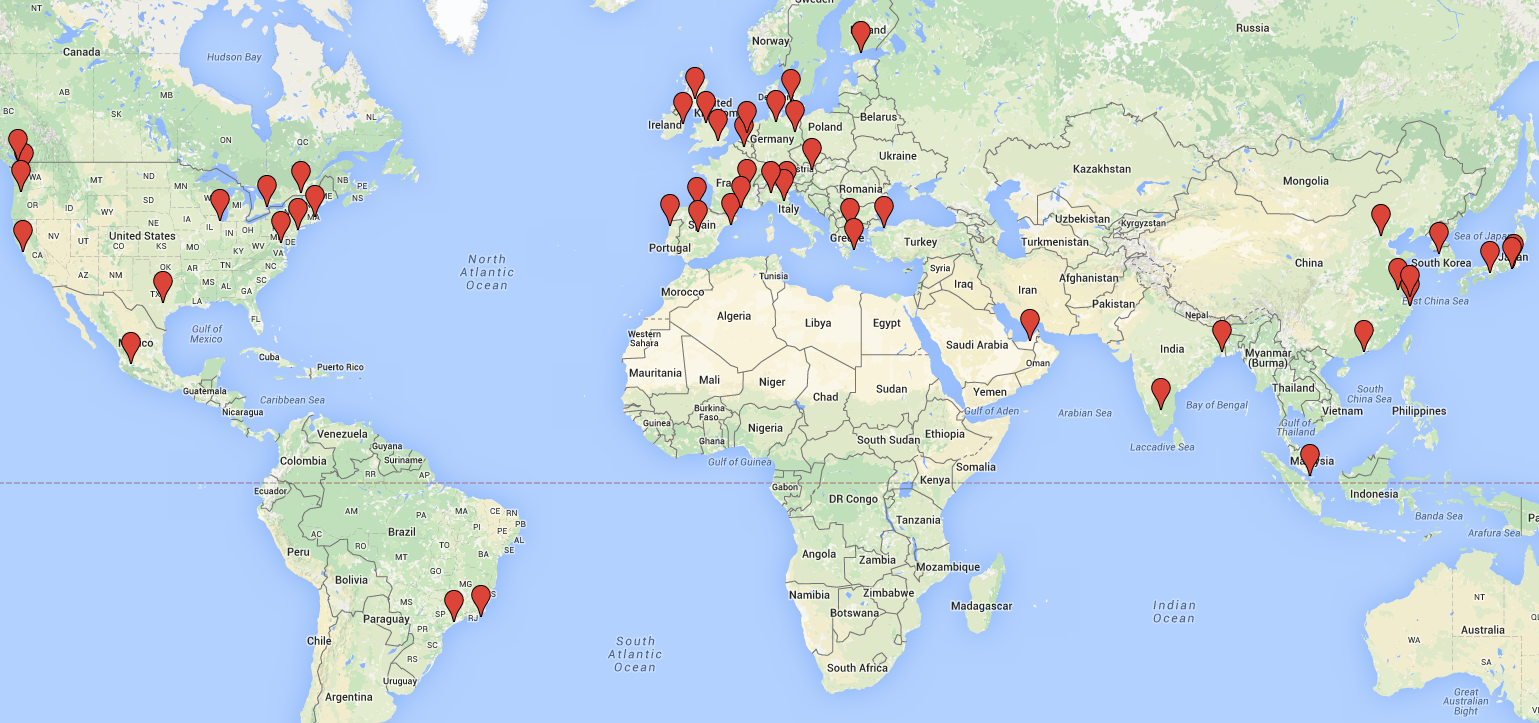}
\caption{Smart Cities initiatives covered in this survey.}
\label{figure:mapa}
\end{figure}

There are multiple smart city initiatives in a variety of countries around the world, with different maturity levels and applications in different domains. Most of the initiatives are in Europe \cite{caragliu2011smart,manville2014mapping}, the USA\footnote{10 Smartest Cities in USA - http://www.fastcoexist.com/3021592/the-10-smartest-cities-in-north-america}, Japan, and South Korea \cite{liu2013smart}. Isolated initiatives exist in countries such as Brazil \cite{fortes2014deployment} and the United Arab Emirates \cite{janajreh2013wind}. Figure \ref{figure:mapa} presents a map with cities that have at least one Smart City project included in this survey. The map shows that most of the projects are located in developed countries, a few in developing countries, and none in underdeveloped countries, where the need for improvements in urban quality of life are most pressing.

\subsection{Enabling Technologies}
\label{subsec:tech-building-blocks}

To answer the question \textit{``What are the main enabling technologies used in state-of-the-art software platforms for Smart Cities?''}, we present the most common enabling technologies that we found in our literature review. We observed four main technologies used by software platforms for Smart Cities: Cyber-Physical Systems, Internet of Things, Big Data, and Cloud Computing. In this section, we give an overview and relate them to Smart City research. These technologies are used later in this paper to group the analyzed platforms and help to understand better the requirements that the platforms must address.

\begin{figure}[!htb]
\centering
\includegraphics[height=7cm]{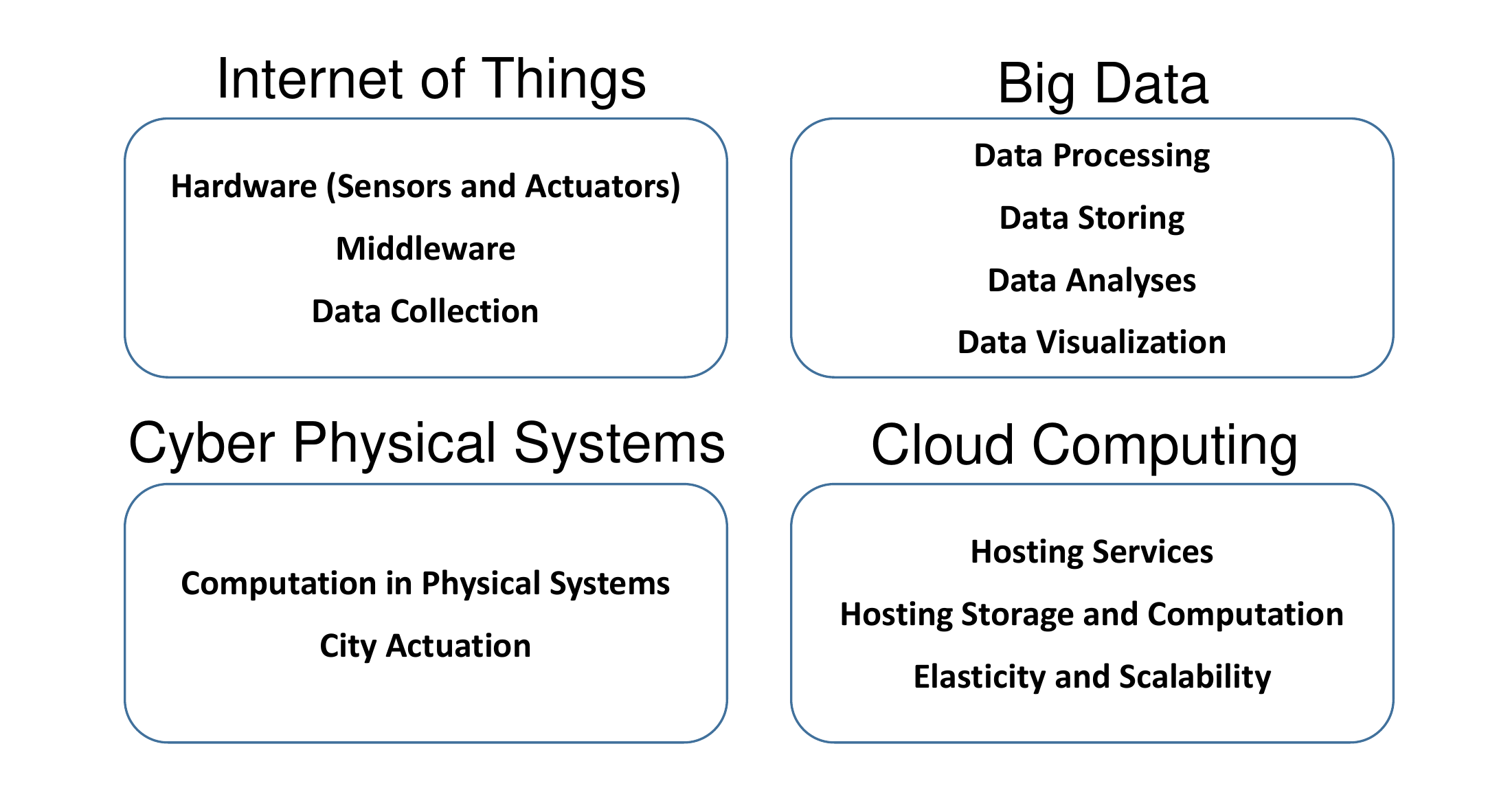}
\caption{Platforms for Smart Cities Enabling Technologies}
\label{figure:blocks}
\end{figure}

Figure \ref{figure:blocks} presents an overview of the four enabling technologies that we found in our survey and examples of how they contribute to a platform for Smart Cities.

\subsubsection{Cyber-Physical Systems}
\label{subsubsec:cyber-physical-systems}

Cyber-Physical Systems (CPS) can be characterized as the use of computation and communication technologies to improve the features of physical systems. Wan et al. \cite{wan2010cpschallenges} define CPS as integration of computation with physical processes. The authors suggest the use of local and remote computational models in networked embedded computers to monitor and control physical processes.

Many real-world applications already leverage CPS \cite{white2010cps}, such as Smart Cities, power grid control systems, and electronic medical devices. However, some authors \cite{wan2010cpschallenges} claim that existing ICT solutions do not support applications with dynamically changing physical contexts. Thus, applying CPS should introduce this requirement to Smart City applications. According to Gurgen et al. \cite{gurgen2013cpssc}, CPS enables applications to become aware of the changes in the physical context adapting their execution according to it.

An example of a Cyber-Physical System related to Smart Cities is WreckWatch \cite{white2010cps}, an application for detecting traffic accidents. This application was developed for smart phones; it reads the device's accelerometer and GPS getting the driver's current speed and acceleration. In case of a strong deceleration, the data is analysed using an accident prediction model and if it indicates an accident, the application generates an alert to a centralized server.

\subsubsection{Internet of Things}
\label{subsubsec:internet-of-things}

Coetzee and Eksteen \cite{coetzee2011iot} define IoT as situations where objects become part of the Internet. According to the authors, the objects have to be uniquely identified, with recognized position and status, and accessible to the network. Gubbi et al. (2013) \cite{gubbi2013internet} define three components in an IoT environment: the hardware, which includes sensors, actuators, and embedded communication hardware; a middleware, which processes and stores data received from the hardware; and a presentation layer, in which users access, manipulate, and visualize data extracted from the hardware. In this sense, this is very similar to what we expect from a platform for Smart Cities. 

The very large number of devices used to collect data from cities forces platforms for Smart Cities to use IoT technologies. The data collected from these devices must be transmitted via interconnected networks so that they can be grouped and processed to provide advanced Smart City services. Zanella et al. \cite{zanella2014iotforsc} present multiple potential uses of the Internet of Things for Smart Cities, e.g., monitoring the health of historical buildings, detecting the load level of waste containers, sensing noise in central areas of the city, observing the conditions of traffic lights, and analyzing the usage of energy in Smart Homes. 
\subsubsection{Big Data}

Most authors consider Big Data as a set of techniques and tools to store and manipulate large data sets whereas conventional technologies, such as relational databases and sequential processing tools, cannot deal with such a vast volume of data. There are four major characteristics of Big Data \cite{chen2014big,demchenko2014defining}: 

\begin{itemize}

\item Volume: the scale of data generated and collected is rapidly increasing, and Big Data tools must deal with this challenge. In Smart Cities, the volume of data will be massive, coming from many data sources distributed across the city. 
\item Variety: data is collected from different sources, and  have structured, semi-structured, or unstructured formats, such as video records, relational databases, and raw texts, respectively. This is important for Smart Cities, since city data is collected from multiple sources, such as surveillance cameras, sensors, and citizen devices. 
\item Velocity: data processing must be fast and, in some cases, real-time, or it may be useless. City infrastructure, operators, and managers need to respond to urban problems, such as traffic jams, accidents, and floods, in short time. 
\item Veracity: because of the large amount of data collected, and the use of multiple data sources, it is important to ensure data quality, because errors in the data or the usage of unreliable sources can compromise its analysis. In cities, incorrect GPS readings, malfunctioning sensors, and malicious users can be sources of poor data. 
\end{itemize}

\begin{figure}[!htb]
\centering
\includegraphics[height=8cm]{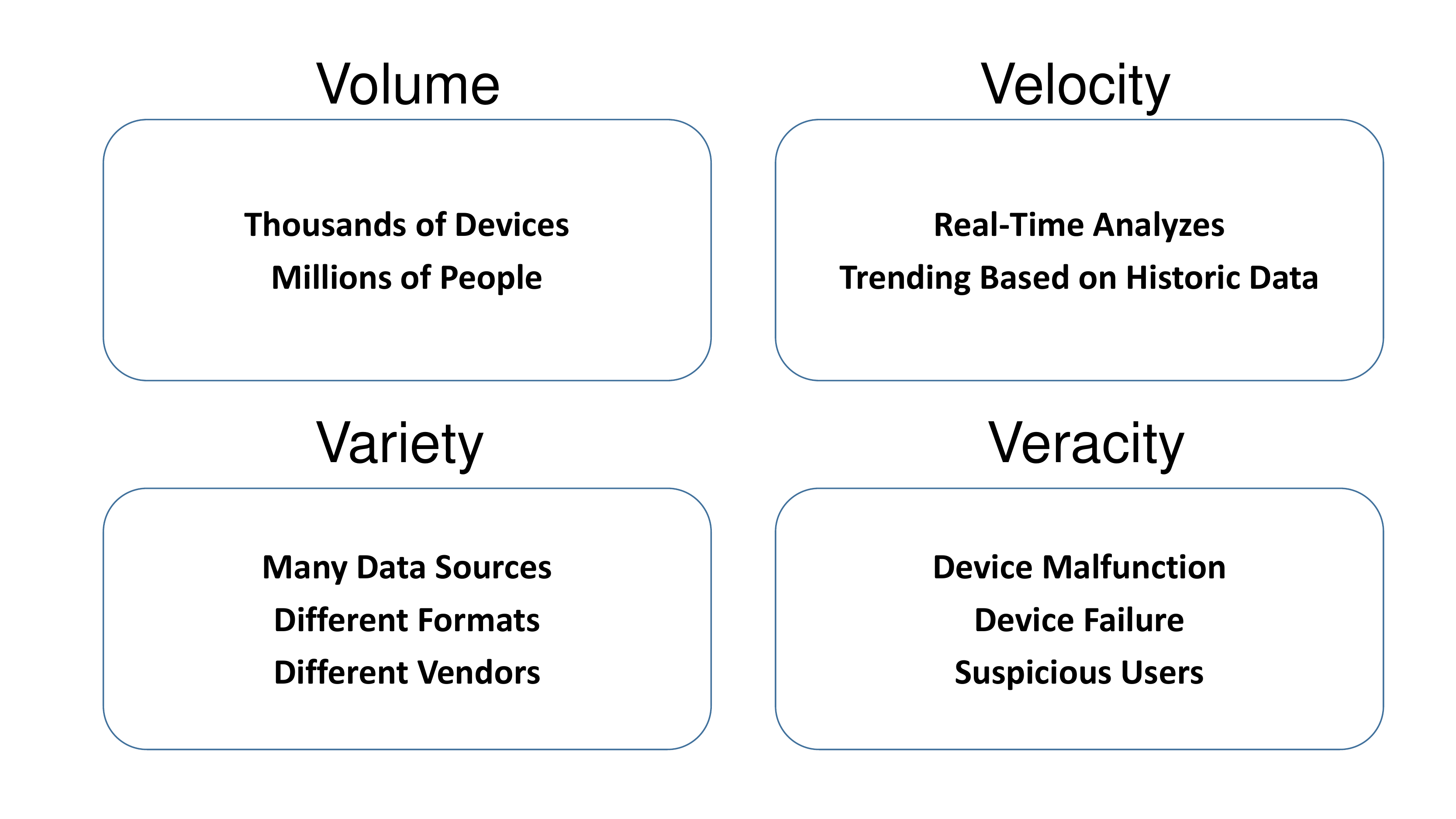}
\caption{4 Vs of Big Data}
\label{figure:bigdata}
\end{figure}

Figure \ref{figure:bigdata} relates the four Vs of Big Data with Smart Cities' needs.

Smart Cities already use Big Data tools to support the amount of data generated from city devices. Sensor networks regularly transmit data about city conditions, such as temperature, air quality, and pluviometry. Citizens generate data using smartphones and social networks, and vehicles continuously send their positions.

Big Data tools are already used by Smart City platforms, including: NoSQL databases \cite{khan2013bigdataanalytics,bain2014sentilo}, such as MongoDB and HBase; parallel data processing tools \cite{parkavi2013hadoopsc,takahashi2012scallop}, such as Apache Hadoop and Apache Spark; real-time data streams processing tools \cite{girtelschmid2013storm}, such as Apache Storm; and visualization tools \cite{khan2013bigdataanalytics}, such as RapidMiner.

Al Nuaimi et al. \cite{applicationsBigData} discuss potential applications of Big Data tools in Smart Cities, such as recognizing traffic patterns and using historic data to locate the causes and avoid traffic jams, facilitating the decisions of city governments using analyses of large data sets, and predicting the use of resources, such as electricity, water, and gas, in different situations using historic and real-time data.

\subsubsection{Cloud Computing}

Cloud Computing offers a very large, elastic, and highly available infrastructure for both data storage and computation, which is essential for complex Smart City systems. In addition, a Smart City environment can be highly dynamic, requiring reconfigurations of the underlying infrastructure, which is also supported by Cloud Computing.

Many authors, such as \cite{distefano2012enabling,aazam2014cloud}, have advocated combining IoT and Cloud Computing, coining the term ``Cloud of Things''. Their idea is to store and process all the data from an IoT network in a cloud computing environment, which is currently used in some Smart City projects \cite{mitton2012cloudsensors,tei2014clout}.

Another concept related to a cloud computing environment in Smart Cities is Software as a Service (SaaS), which provides the sensor data with a cloud computing infrastructure. The work of Perera et al. \cite{perera2014iot} extends this concept, using the term ``Sensing as a Service''. The ClouT platform, presented in \cite{tei2014clout}, also uses the concept of software services and defines the terms City Application Software as a Service (CSaaS) and City Platform as a Service (CPaaS).

Some authors relate the use of Cloud Computing, Big Data, and IoT \cite{chen2014big,aazam2014cloud}, because a cloud environment is an ideal infrastructure to store data and execute services. Hence, the data generated from an IoT middleware can be stored and processed in a cloud environment using Big Data tools. This synergistic combination helps to support important non-functional requirements such as scalability, elasticity, and security.

\section{Platforms for Smart Cities}
\label{sec:software-platforms}

We describe here various platforms for Smart Cities presented in the literature. All platforms use at least one of the enabling technologies discussed in Section \ref{subsec:tech-building-blocks}.

\begin{figure}[!htb]
\centering
\includegraphics[scale=0.5]{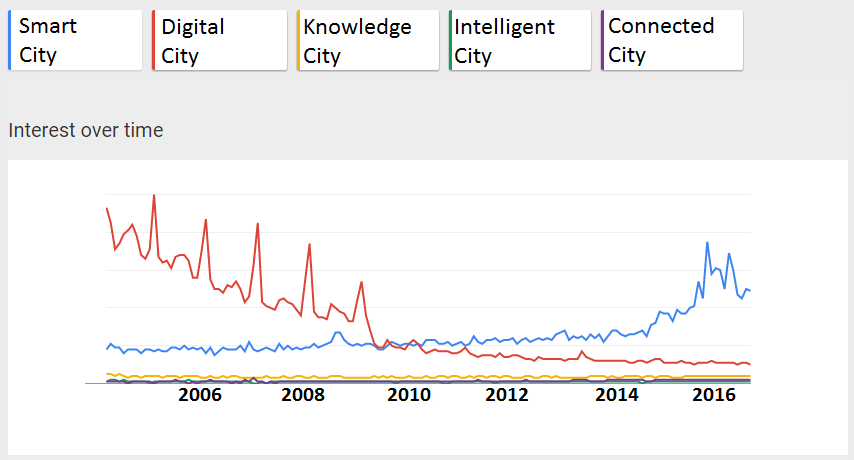}
\caption{Trends in Smart City related terms}
\label{figura:trends}
\end{figure}

To find these studies, we used the following query string: \textsf{(``Smart City'' or ``Smart Cities'') and (Platform or Middleware or Architecture)}. After analyzing the query results, we focused our study on 47 papers describing Smart City platforms and applications. We did not include in our search other terms that are more rarely used to describe the application of ICT in cities, such as ``Knowledge City", ``Intelligent City", and ``Connected City".  Figure  \ref{figura:trends} illustrates the use of these expressions in recent years using Google Trends.

Since the expression ``Digital City" is still used, we analyzed the definition of this expression and the differences with ``Smart Cities". We found that, normally, the description of a digital city relates to the use of digital technologies in a city, but not with the goal of making smart services and improving the city's overall infrastructure. In a digital city, the integration of the multiple systems is not at stake. The differences between these two concepts are discussed by Cocchia \cite{cocchia2014systematic} and by Yin et al. \cite{yin2015surveySmartCities}.

The next subsection describes existing platforms, developed as research projects with different approaches. Subsection \ref{subsec:systems} shows systems developed using these platforms. Finally, in Subsection \ref{subsec:requirements}, we present a set of functional and non-functional requirements extracted from our analysis of the platforms and systems described in the previous subsections.

\subsection{Platform Categories} 
\label{subsec:categories}

\begin{figure}[!htb]
\centering
\includegraphics[scale=0.4]{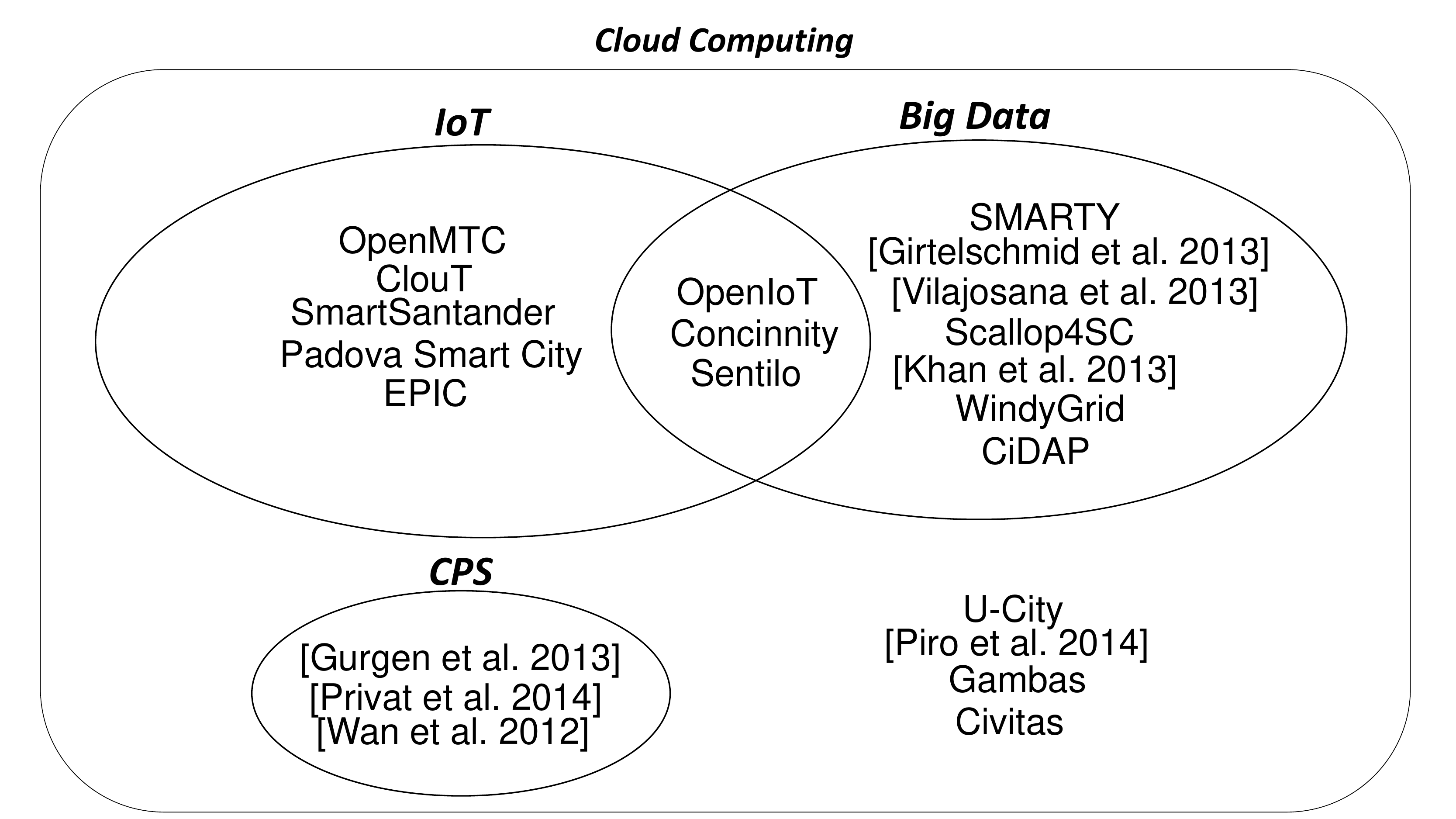}
\caption{Use of Enabling Technologies by Smart City Platforms}
\label{figura:overview}
\end{figure}

To facilitate the presentation, we divided the platforms into five categories, according to the enabling technologies that each platform uses. Figure \ref{figura:overview} presents an overview of the platforms for Smart Cities that we analyzed. In this figure, we can observe that most platforms use Cloud Computing. Almost all of them use at least one more enabling technology, more commonly IoT and Big Data.

\subsubsection{Internet of Things and Cloud Computing}
\label{subsubsec:iot-and-cloud}

In this section, we present the platforms that use both IoT and Cloud Computing as enabling technologies. 

SmartSantander is an experimental infrastructure to support the development and deployment of Smart City applications and services \cite{sanches2014smartsantander}. The project is centered in Santander, Spain, with smaller facilities in other European cities. The platform processes a large variety of information, including data about traffic conditions, temperature, $CO_2$ emissions, humidity, and luminosity. Currently, the project has implanted more than 20,000 sensors in the city.

Padova Smart City \cite{zanella2014iotforsc} uses IoT to create a sensor network in the city of Padova, Italy. Using more than three hundred sensors, the platform collects environmental data, such as CO2 emissions and air temperature, and monitors street lights. A feature highlighted in this platform is the use of common protocols and data formats to allow interoperability among multiple city systems.

The European Platform for Intelligent Cities (EPIC) project \cite{ballon2011there} proposes a complete IoT Middleware to facilitate the use and management of the Wireless Sensor Network (WSN). This middleware aims to deal with the heterogeneity, interoperability, scalability, extensibility, and configurability problems in a WSN.

ClouT \cite{tei2014clout} proposes a two-layer architecture to collect data from the WSN and manage the sensors and actuators in the city network \cite{cloutPlatform}. The first layer is the Sensors and Actuators Layer, which handles data from the WSN. The second layer, the IoT Kernel Layer, manages and monitors the sensors and actuators network.

OpenMTC \cite{elmangoush2013openmtc} (Open Machine Type Communications) is a Machine-To-Machine (M2M) based communication platform for Smart Cities. Its goal is to enable efficient communication among a large number of devices, associating them with multiple services. To achieve this, the platform supports standard interfaces to various types of devices, data/event processing methods to achieve real-time performance, and easy application development, providing a software development kit.

The analysis of the platforms aforementioned led to the identification of four major functional requirements: management of a WSN,  management of the data collected from the city, management of services and applications, and an infrastructure to make the data from the platform available to city applications. This analysis also led to the identification of five non-functional requirements: adaptation, interoperability, scalability, extensibility, and configurability.

We identify two weak points of these platforms: (1) the lack of pre-processing components to verify the integrity of the data collected from the city and make small transformations of the data, such as aggregations, and (2) most of the platforms do not include a discussion about security concerns.

\subsubsection{Internet of Things, Cloud Computing, and Big Data}
\label{subsubsec:iot-cloud-big-data}

In this subsection, we present platforms that use IoT, Cloud Computing, and Big Data as enabling technologies.

OpenIoT\footnote{OpenIoT - https://github.com/OpenIotOrg/openiot} is an open source middleware for the development of IoT-based applications. It has an API to manage the WSN, and a directory service to dynamically discover the sensors deployed in the city; it also has a layer for service definition and access. Big Data tools are used to store and analyze the data from the platform. A Smart City project called Vital \cite{petrolo2014cloudofthings} builds on this platform and uses the term ''Cloud of Things" to refer to the use of Cloud Computing and IoT. 

The Concinnity project provides a platform for managing data and applications following the PaaS model \cite{chao2014concinnity}, with which its authors built Big Sensor Data Applications. However, this platform focuses on multiple data sources such as the WSN, social networks, and data from platform users. It also includes a service directory where developers can find and publish services facilitating its reuse.

OpenIoT and Concinnity, offer developers tools to implement applications directly on the platform. OpenIoT allows the mash-up of the services defined in the platform and automatically creates a visual interface for end-users. Concinnity provides a set of development tools, such as a Workflow Editor and Engine, a Service Publisher, and an Application Editor. 

Sentilo \cite{bain2014sentilo} is platform that deals with the management of sensors and actuators, designed for Smart Cities that looks for openness and interoperability. Sentilo uses IoT concepts to control the WSN, and Cloud Computing to share data with the applications. Big Data tools are mainly used to collect and store data from sensors, ensuring platform scalability. The Sentilo project was originally designed to be deployed in the city of Barcelona; after its deployment, the City released the code under the LGPL and EUPL open source licenses.

The main functional requirements identified for this group of platforms were: management of a WSN, management of data life cycle (collect, store, process), making the data from the platform publicly available, a service directory for application developers, and tools for application development. As non-functional requirements, we identified: interoperability and scalability.

A weak point of these platforms is the lack of streams processing tools to analyze real-time data from the city, an important requirement for many Smart City applications. Another problem is that most of the platforms do not support the customization of services with citizen data. In spite of the privacy problems, offering context-aware, customized services to the citizens is highly desirable. 

\subsubsection{Cloud Computing and Big Data}
\label{subsubsec:cloud-big-data}

In this subsection, we present platforms that use Cloud Computing and Big Data as enabling technologies. 

Vilajosana et al. \cite{vilajosana2013bootstrapping} present a platform for Smart Cities based on Cloud Computing and Big Data, whose main components are data management and service hosting. It includes an Open Data API allowing third-party applications to access the data stored on the platform. Big Data tools are used to collect data streams and analyze data, such as prediction and inference. 

Scallop4SC (SCALable LOgging Platform for Smart City) \cite{takahashi2012scallop,yamamoto2014using} uses Big Data to process a large volume of data gathered from smart buildings. The platform uses information about the building, such as water and energy consumption, temperature, air humidity, and the amount of garbage generated. Periodically, the buildings send data to the platform for processing. The objective is to analyze smart building data, for which it uses the MapReduce algorithm.

CiDAP \cite{cheng2015bigdataplat} is a big data analytics platform deployed into the SmartSantander testbed. The platform uses data collected from SmartSantander and analyzes it to understand the behavior of the city. The main components of this platform are: the agents, which collect data from the SmartSantander platform; the Big Data repository for storing the data; the Big Data processing for intensive data processing and analytics; and a CityModel server, responsible for interfacing with  external applications. This platform uses Apache Spark \cite{zaharia2010spark} to process the data.

\cite{Khan2015cloudbasedplatform} propose a Smart City architecture based on Big Data to achieve the necessary availability and scalability required for a Smart Cities platform. The architecture has three layers: one to collect, analyze, and filter data; another to map and aggregate data to make it semantically relevant; and a third layer where users can browse and recover the data processed from the other two layers. The implementation of the architecture uses only open source projects, and the authors have presented tools for all layers \cite{khan2013bigdataanalytics}.

WindyGrid \cite{thornton13windygrid}, an initiative of the City of Chicago, is a platform for Smart Cities, whose objective is to present real-time and historical data with a unified view of city operations. Big Data technologies, such as the MongoDB NoSQL database and parallel data processors, were used to develop the platform.

SMARTY \cite{anastasi2013urban} is a project aimed at providing tools and services for mobility and flexible city transport systems. Its software platform collects data from multiple sources, such as traffic flow, user location, transport service delays, and parking availability. A network of low-cost sensors collects data from the city and social networks are continuously monitored to get data from citizens. The platform processes the massive amount of data generated by the city with data mining techniques, such as classification, regression, and clustering.

The platform proposed by Girtelschmid et al. \cite{girtelschmid2013storm} uses semantic technologies to create a platform for Smart Cities, adding flexibility in system configuration and adaptation. However, to overcome the performance bottlenecks normally associated with ontology repositories and reasoning tools, the authors combine their semantic techniques with Big Data processing methods.

The main functional requirements identified for this group of platforms were: data management, such as collecting, analyzing, and visualizing data; large scale data processing, such as batch and real-time processing; and the use of semantic techniques combined with Big Data. As non-functional requirements, we identified scalability and adaptation.

Most of the platforms in this section do not have an IoT layer, and do not indicate how the data is collected from the city; the exception is CiDAP, which uses the SmartSantander testbed as an IoT middleware. Another drawback is that most of the platforms do not include a discussion about security concerns.

\subsubsection{Cloud Computing}
\label{subsubsec:cloud}

In this subsection, we present platforms that use only Cloud Computing as an enabling technology. 

Piro et al. \cite{piro2014information} present a two-layered service platform for the creation of Smart City applications. The first is a low-level layer that controls the communication among the city WSN devices. The second layer collects the data from the devices and provides services for the development of applications that use the data from the city.

U-City \cite{woo2010ucity} is a platform for the creation of smart ubiquitous cities. The platform offers several service management features, such as autonomic service discovery, service deployment, and context-aware service execution. It also offers predefined services such as an inference engine, a context-aware data service, and a portal for the management of the platform.

Gambas, a middleware for the development of Smart City applications \cite{apolinarski2014gambas}, supports data acquisition, distribution, and integration. The platform also provides an application runtime to facilitate the development and deployment of services using city data and a service registry. The middleware supports context-awareness, so that Smart City services can adapt to the citizen situation, behavior, and intent. All communication in the platform is encrypted to ensure citizen's privacy and security.

Civitas \cite{villanueva2013civitas} is a middleware to support the development of Smart Cities services. It is used to facilitate the development and deployment of Smart City applications, and to avoid the emergence of ``information islands'' \cite{junping2010digital}, i.e., disconnected applications that do not share relevant information. Citizens connect to the middleware via a special device called the Civitas Plug, which ensures the privacy and the security. The middleware has two main design principles to facilitate the application integration: \emph{Everything is a Software Object}, which promotes the consistency of the software design and reusability of the middleware; and \emph{Independence of the City Layout}, meaning that city services should not work with just one city layout.

The main functional requirements identified for this group of platforms were: service management and data management. As non-functional requirements, we identified: security, privacy, and context awareness.

A drawback of the platforms presented in this section is that none of them use known frameworks to implement components, such as the inference engine and processing tools, which might make difficult the maintenance of the platform. Another problem is that the platforms do not describe a mechanism to allow external access to the platform data.

\subsubsection{Cloud Computing and Cyber-Physical Systems}
\label{subsubsec:cloud-cyber-systems}

In this section, we present platforms that use Cloud Computing and Cyber-Physical Systems (CPS) as enabling technologies. 

Gurgen et al. \cite{gurgen2013cpssc} present a middleware for Smart Cities autonomic services, which includes many ``self-'' properties, such as self-organization, self-optimization, self-configuration, self-protection, self-healing, self-discovery, and self-description. They justify using cloud computing to provide scalability, reliability, and elasticity to the platform. This platform provides application developers with the contexts of individual users and the city.

Privat et al. \cite{privat2014towards} propose another CPS-based platform, whose main characteristic is self-configuration and self-adaptation capabilities in smart environments, including Smart Cities. This platform provides a shared distributed software infrastructure that collects data and reacts to changes in the environment.

Wan et al. \cite{wan2012m2m} propose an event-based CPS platform, which uses an event manager to manage and generate cooperation among M2M components. This platform provides data and services to third-party applications through a publish/subscribe module. The platform also enables the design of event processing flows to manage the mission-critical wireless messages.

The main functional requirements identified for this group of platforms were: autonomic reaction to changes in the city environment, communication among city devices, and a publish/subscribe mechanism for applications to communicate with the platform. As non-functional requirements, we identified: configurability, adaptation, and context awareness.

The platforms of this section focus on the deployment, configuration, and execution of CPS devices in the city, but they lack important requirements, such as the monitoring and publication of the data from the devices. They also do not describe any mechanism to verify the data collected from the city, discarding inconsistencies.

\subsection{Systems}
\label{subsec:systems}

In this subsection, we illustrate applications built on top of the platforms presented in the previous section. Table \ref{tab:systems} presents an overview of the domains of the analyzed systems.

\begin{table}[!htbp]
\label{table:systems_table}
\tbl{Domains of Smart City Systems\label{tab:systems}}{%
\begin{tabular}{|l|l|l|l|l|l|l|l|l|l|l|}
\hline

& \begin{sideways}City Sensing\end{sideways}
& \begin{sideways}Traffic Control\end{sideways} 
& \begin{sideways}Air Pollution\end{sideways}
& \begin{sideways}City Dashboard\end{sideways}
& \begin{sideways}Health Care\end{sideways}\
& \begin{sideways}Safety\end{sideways}
& \begin{sideways}Disaster Prevention\end{sideways}
& \begin{sideways}Energy Management\end{sideways}
& \begin{sideways}Waste Management\end{sideways}
\\
\hline
GAMBAS 
&   & X &   &   &   &   &   &   & \\
\hline
SmartSantander
& X & X & X &   &   &   &   &   & \\
\hline
Padova Smart City 
& X &   &   & X &   &   &   &   &  \\
\hline
OpenIoT
& X &   &   &   &   &   &   &   & X\\
\hline
WindyGrid  
& X & X & X &   & X & X &   &   & \\
\hline
ClouT
& X & X &   &   & X & X & X &   & \\
\hline
Scallop4SC
&   &   &   &   &   &   &   & X & \\
\hline
Number of Instances
& 5 & 4 & 2 & 1 & 2 & 2 & 1 & 1 & 1 \\

\hline
\end{tabular}}
\end{table}%

The GAMBAS middleware was used to develop two applications for the public transportation system in Madrid, Spain. Foell et al. \cite{Foell2014urbanbus} present a context-aware urban bus navigator to help travelers find the best buses for their trips. Handte et al. \cite{handte2014crowd} describe a system that estimates the number of passengers of city buses using smartphone sensing.

SEN2SOC \cite{vakali2014datastream} is a system deployed on the SmartSantander platform that uses data streams from the city (e.g., sensor data) and social networks (e.g., Twitter) to create Smart City applications. Two examples of applications are: capturing the emotional state of city inhabitants, and visualizing the air pollution in the city. Also in SmartSantander, Vlahogianni et al. \cite{vlahogianni2014exploiting} present an application to predict the utilization of city parking lots.

Two applications were developed using the Padova Smart City platform. Bui and Zorzi \cite{bui2011health} present a health care system whose main features are monitoring conditions of patients, sending their data directly to doctors, and calling emergency services if the patient has an urgent problem. Bressan et al. \cite{5622015} present a monitoring application to manage and collect data from all the light posts connected to the platform.

Anagnostopoulos et al. \cite{AnagnostopoulosWasteOpenIoT} present a waste management system implemented using the OpenIoT platform. It presents four models to prioritize critical trash bins, such as bins close to schools, hospitals, and gas stations. The system was used to compare the four models according to the amount of CO2 emitted and the distance traveled by trash trucks.

The WindyGrid platform \cite{Rutkin201424} provides three main systems to the city of Chicago: Situational Awareness and Incident Monitoring, to monitor and act on problems that are occurring in the city; Historical Data Analyses, to predict the behavior of city systems, such as traffic and health care; and Advanced Real-Time Analytics, to analyze the current situation of city systems. Some examples of the data used in these systems are: log of emergency (911) calls, traffic conditions, public buildings information, and surveillance cameras.

Galache et al. \cite{cloutPlatform} present four systems developed using the ClouT platform: an alert service to warn citizens about earthquakes in Fujisawa, Japan; a civil protection system, which warns the population about environmental risks such as storms and earthquakes in Genova, Italy; a system to help elderly people find healthy activities in the city of Mitaka, Japan; and a sensing application to notify people about events in Santander, Spain, such as cultural acts and traffic accidents.

Yamamoto et al. \cite{yamamoto2014using} present two systems developed for the Scallop4SC platform, both in the energy management domain. The first system offers a tool for the visualization of household energy consumption, which analyzes data at different levels, such as state, city, and neighborhood. The other system is a wasteful energy detection service that is available for smart homes.

The analyzed applications show that the most explored domains in the literature are traffic, with applications to monitor the streets or help citizens to use the public transport, and city sensing, capturing data from the city using sensors such as air pollution and temperature. Most of the applications are developed external to the platform, using only one or more platform services.

\subsection{Requirements for Smart City Software Platforms}
\label{subsec:requirements}

To answer the second research question \textit{``What are the requirements that a software platform for Smart Cities should meet?''}, in this section we analyze the functional and non-functional requirements extracted from the analyzed platforms.

We assume that a platform implements a requirement if the literature describing it explicitly states so, or if the platform has a component or module that clearly fulfills that requirement.

\subsubsection{Functional Requirements}
\label{subsubsec:functional}

The main goal of a platform for Smart Cities is to facilitate the development of Smart City applications. Towards this aim, most of the analyzed platforms implement requirements for collecting data from the city, managing and sharing data, and providing tools to facilitate the development of Smart City applications. Table \ref{tab:four} presents an overview of the functional requirements for Smart City platforms, which we describe in the following:

\begin{table}[!htbp]
\tbl{Functional requirements for Smart City platforms\label{tab:four}}{%
\label{tab:functional_table}
\begin{tabular}{|l|c|c|c|c|c|c|c|c|}
\hline

& \begin{sideways}Data Management\end{sideways} 
& \begin{sideways}Application Run-time\end{sideways}
& \begin{sideways}WSN Management\end{sideways}
& \begin{sideways}Data Processing\end{sideways}
& \begin{sideways}External Data Access\end{sideways}
& \begin{sideways}Service Management\end{sideways}
& \begin{sideways}Software Engineering Tools\end{sideways}
& \begin{sideways}Definition of a City Model\end{sideways}\\\hline

SmartSantander    
& X & X & X &   & X &   &   &  \\\hline
OpenIoT           
& X & X & X & X &   & X & X & \\\hline
Concinnity        
& X & X &   & X & X & X & X & \\\hline
Civitas           
& X &   &   & X &   & X &   & \\\hline
Gambas            
& X & X &   &   & X & X &   & X \\\hline
\cite{khan2013bigdataanalytics}    
& X &   &   & X & X & X & X & \\\hline
\cite{girtelschmid2013storm}  
&   &   &   & X & X &   &   & \\\hline
Scallop4SC        
& X &   &   & X & X &   &   & \\\hline
OpenMTC           
&   &   &   &   & X & X & X &  \\\hline
\cite{wan2012m2m}    
& X &   &   & X &   & X &   &   \\\hline
\cite{piro2014information} 
&   &   &   &   &   & X &   & \\\hline
\cite{gurgen2013cpssc}    
& X & X & X & X & X & X &   & \\\hline
\cite{vilajosana2013bootstrapping}  
& X &   & X & X & X & X &   & \\\hline
ClouT            
& X & X & X &   & X &   &   & \\\hline
Padova Smart City 
& X &   & X & X & X &   &   &  \\\hline
U-City            
& X & X &   & X & X & X &   &  \\\hline
Sentilo           
& X &   & X &   & X &   &   & \\\hline
WindyGrid         
& X &   &   &   &   & X &   & \\\hline
EPIC         
& X &   & X &   & X & X &   & \\\hline
\cite{privat2014towards} 
&   &   &   &   &   &   &   & X \\\hline
SMARTY 
& X &   & X & X & X & X &   & \\\hline
CiDAP  
& X &   &   & X & X &   &   & X \\\hline
Number of Instances  
& 18 & 7 & 9 & 13 & 16 & 14 & 4 & 3 \\\hline

\end{tabular}}
\end{table}%

\begin{itemize}

\item \textbf{Data Management:} This is a requirement implemented by most of the platforms for Smart Cities, which includes collection, storage, analysis, and visualization of city data. The analyzed platforms use different techniques for this requirement, such as relational databases \cite{munoz2011forefront,woo2010ucity}, big data tools \cite{thornton13windygrid,cheng2015bigdataplat}, and customized tools implemented by the platform development team \cite{chao2014concinnity}.

\item \textbf{Applications Run-time:} Some platforms focus on managing the execution of their applications. The goal is to facilitate the deployment and integration of such applications. Some platforms provide a complete environment for developers to deploy their applications \cite{apolinarski2014gambas}; others offer an execution run-time service for applications developed with tools the platform provides \cite{petrolo2014cloudofthings,chao2014concinnity}. 

\item \textbf{WSN Management:} Many of the analyzed platforms have a Wireless Sensor Network (WSN) management layer to control and monitor the devices deployed in the city. Most of these platforms use IoT concepts to organize and manage the WSN \cite{munoz2011forefront,tei2014clout}. Other platforms \cite{bain2014sentilo} do not explicitly mention this, but indeed have a software layer to manage the city network devices. Some platforms include features to manage all the device activities, such as adding, removing, and monitoring the sensors and actuators. Two platforms describe a WSN deployed in a city: Padova Smart City \cite{zanella2014iotforsc}, with 3000 sensors, and SmartSantander \cite{sanches2014smartsantander}, with more than 20000 sensors.

\item \textbf{Data Processing:} Some platforms use specific processing components, such as inference engines \cite{woo2010ucity}, workflow processing \cite{chao2014concinnity}, and big data processing tools \cite{takahashi2012scallop}. These components process large data sets, and their main purpose is to analyze, verify, aggregate, and filter the data from the city. In addition, some platforms \cite{girtelschmid2013storm,cheng2015bigdataplat} make real-time analyses of data streams.

\item \textbf{External Data Access:} Almost all platforms describe an interface for external applications to access the platform data. The most common approach is an API to allow access to the data generated in the city. Some platforms use REST \cite{munoz2011forefront,elmangoush2013openmtc}, others use cloud computing concepts to provide the city data as a service \cite{ballon2011there}, and one proposes an open data platform \cite{zanella2014iotforsc}. Also, a platform \cite{gurgen2013cpssc} uses the publish/subscribe paradigm to make the data and services available to applications. 

\item \textbf{Service Management:} Most of the analyzed platforms adopt a Service-Oriented Architecture, in which the platform functionalities are offered by services \cite{issarny2011jisa}. Some of them use services to provide features to applications, such as access to raw sensors data \cite{petrolo2014cloudofthings} and analyzed data \cite{zanella2014iotforsc}, and workflow engines \cite{chao2014concinnity}. Others enable developers to deploy services on the platform and make them available to other applications \cite{apolinarski2014gambas,piro2014information}. Some platforms also use service compositions and choreographies \cite{issarny2011jisa} to create new services or applications \cite{woo2010ucity,piro2014information}. 

\item \textbf{Software Engineering Tools:} Some platforms provide a set of tools for the development and maintenance of services and applications. For describing and implementing applications, some platforms create visual interfaces \cite{petrolo2014cloudofthings}. Other platforms provide workflow design tools \cite{chao2014concinnity} to define data or service flows and create Smart City applications. Moreover, some platforms \cite{khan2013bigdataanalytics} use analytics and reporting tools to facilitate the development of data visualization and reports, and two platforms describe the use of a Smart City application SDK \cite{elmangoush2013openmtc,apolinarski2014gambas}. 

\item \textbf{Definition of a City Model:} Some platforms provide a model of the city to facilitate the manipulation and understanding of the platform data, and to facilitate the integration of the collected data. For example, in Cheng et al. \cite{cheng2015bigdataplat}, the city model is used to allow queries in the data from the city sensor network. Privat et al. \cite{privat2014towards} use a finite-state model to represent the possible city data flows.

\end{itemize}

Based on the functional requirements aforementioned, we can observe that the main platforms activities aim to control the city data life cycle: (1) Collecting the data with a WSN, (2) Managing the data in the platform, (3) Processing the data using city models, and (4) Sharing the raw and processed data allowing external access. These activities are highly related to the enabling technologies, such as IoT with the WSN management, Data Management and Processing with Big Data, and Service Management with Cloud Computing.

\subsubsection{Non-Functional Requirements}
\label{subsubsec:requirements}

Most of the non-functional requirements of Smart City platforms are related to large, heterogeneous distributed systems, such as scalability, adaptation, and interoperability. Other non-functional requirements are related to the manipulation of critical and personal data from citizens, such as security and privacy. Table \ref{tab:nonfuncionalreq} presents an overview of the non-functional requirements for Smart City platforms, which we describe in the following.

\begin{table}[!htbp]
\tbl{Non-Functional requirements for Smart City platforms\label{tab:nonfuncionalreq}}{%
\label{tab:non-functional_table}
\begin{tabular}{|l|c|c|c|c|c|c|c|c|}
\hline

& \begin{sideways}Interoperability\end{sideways} 
& \begin{sideways}Scalability\end{sideways}
& \begin{sideways}Security\end{sideways}
& \begin{sideways}Privacy\end{sideways}
& \begin{sideways}Context Awareness\end{sideways}
& \begin{sideways}Adaptation\end{sideways}
& \begin{sideways}Extensibility\end{sideways}
& \begin{sideways}Configurability\end{sideways}\\\hline

SmartSantander    
&   & X & X & X &   &   &   &  \\\hline
OpenIoT           
& X &   & X &   &   &   &   & X \\\hline
Concinnity        
& X &   & X & X &   &   &   & X \\\hline
Civitas           
& X &   & X & X &   &   &   & \\\hline
Gambas            
&   &   & X & X & X &   &   &   \\\hline
\cite{khan2013bigdataanalytics}    
&   & X &   &   & X &   & X & \\\hline
\cite{girtelschmid2013storm}  
& X & X &   &   & X & X &   & \\\hline
Scallop4SC        
&   & X &   &   &   &   & X & \\\hline
OpenMTC           
& X &   &   &   &   &   &   &  \\\hline
\cite{wan2012m2m}    
& X &   &   &   &   & X &   & X  \\\hline
\cite{piro2014information} 
&   &   & X &   &   &   &   & \\\hline
\cite{gurgen2013cpssc}    
& X &   &   &   & X & X &   & \\\hline
\cite{vilajosana2013bootstrapping}  
& X &   &   &   &   &   &   & \\\hline
ClouT            
& X &   & X &   &   &   &   & \\\hline
Padova Smart City 
& X & X &   &   &   &   &   &  \\\hline
U-City            
&   &   & X &   &   &   &   & X \\\hline
Sentilo           
&   & X & X &   &   &   & X & \\\hline
WindyGrid         
&   &   & X & X  &   &   &   & \\\hline
EPIC         
& X & X &   &   &   &   &   & \\\hline
\cite{privat2014towards} 
&   &   &   &   & X & X &   & X \\\hline
SMARTY 
& X &   &   &   & X &   &   & \\\hline
CiDAP  
& X & X &   &   & X & X & X &   \\\hline
Number of Instances  
& 13 & 8 & 10 & 5 & 7 & 5 & 4 & 5 \\\hline
\end{tabular}}
\end{table}%

\begin{itemize}

\item \textbf{Interoperability:} Different devices, systems, applications, and platforms compose a Smart City environment, and all these components must operate in an integrated fashion; for example, sensors from multiple vendors, systems implemented in different languages, platforms that share data and users, and legacy systems that have to communicate with the new platforms. Previous work in the field adopted several techniques to handle this requirement: interoperable objects \cite{villanueva2013civitas}, adopting generic and standard interfaces \cite{gurgen2013cpssc}, applying Semantic Web to integrate all platform components \cite{girtelschmid2013storm}, and using a naming mechanism \cite{cheng2015bigdataplat} to recognize different devices or data sources.

\item \textbf{Scalability:} The amount of users, data, and services of a Smart City platform will be massive, and can increase over time. For example, in the SmartSantander testbed, there were more than 20,000 sensors, in a city of 178,000 inhabitants collecting a large amount of city data \cite{sanches2014smartsantander}; CiDAP collected more than 50 GBs of data in three months \cite{cheng2015bigdataplat}. This non-functional requirement is relevant to many functional requirements, such as WSN management \cite{ballon2011there}, data management \cite{takahashi2012scallop}, and service management \cite{bain2014sentilo}. 

\item \textbf{Security:} Malicious users can make fraudulent use of services and data provided by the platform. Many platforms have a component or describe mechanisms to handle security, avoiding attacks to the city infrastructure and information theft \cite{piro2014information,munoz2011forefront,petrolo2014cloudofthings}. 

\item \textbf{Privacy:} A Smart City platform collects and manipulates several citizen-sensitive data, such as medical records, user localization, and consuming habits. The challenge is to use these data while hiding, or to avoid saving identifiable information. Some of the strategies used to achieve this requirement are cryptography \cite{apolinarski2014gambas}, tokens to control the access to the data that users can manipulate \cite{villanueva2013civitas}, and anonymization \cite{mylonas2015smartphonesantander}. 

\item \textbf{Context Awareness:} As the city and user situation can change over time, many applications and services can provide better results using contextual information. Some platforms use information from users \cite{apolinarski2014gambas,privat2014towards}, such as location, activity, and language. Other platforms use information from the city \cite{khan2013bigdataanalytics,cheng2015bigdataplat}, such as traffic conditions, climate, and air quality. Examples of context use are: displaying a different language in an application to a tourist, and changing the route of a user avoiding polluted areas. 

\item \textbf{Adaptation:} Related to context awareness, many platforms adapt their behavior based on context in order to achieve fault-tolerance, choose a closer server to improve efficiency, decide for batch or real-time processing, and adapt data from multiple data sources. This requirement is most used in platforms that use CPS as enabling technology \cite{privat2014towards,wan2012m2m}, but other concepts are used to meet this requirement as well, such as semantic technologies \cite{girtelschmid2013storm}. 

\item \textbf{Extensibility:} The capability to add services, components, and applications to the platform is important to assure that it meets evolving system requirements and user needs. Mun\~oz et al. \cite{munoz2011forefront} state that easy extensibility is valuable because one cannot know what services a city will need. Scallop4SC \cite{takahashi2012scallop} uses materialized views that developers extend to implement their applications. Some platforms \cite{khan2013bigdataanalytics,bain2014sentilo} employ only open source tools, facilitating the platform's extensibility. CiDAP \cite{cheng2015bigdataplat} offers extensibility to enable the use of the platform in cities of different scales. 

\item \textbf{Configurability:} A Smart City platform has many configuration options and parameters that define its behavior at execution time, such as defining pollution and congestion thresholds and the priority of services. Thus, it is important to allow (re)configuration of the many variables of the platform. Two platforms \cite{wan2012m2m,privat2014towards} highlighted the importance of self-configurability capacities, because of the huge amount of configurations needed in a Smart City platform. Other platforms \cite{woo2010ucity,kim2014openiot} provide a portal to centralize the configurations. 

\end{itemize}

Based on the non-functional requirements aforementioned, we can observe that some of them are very important to many functional requirements: such as Scalability, which is valuable to the WSN and Data Management; Security and Privacy, which are important to all data requirements; Extensibility, which is required to the Service Management; and Configurability, which is important to all the functional requirements. The non-functional requirements are very similar to the challenges and open research problems that we present in the next section.

\section{Challenges and Open Research Problems}
\label{sec:challenges}

To answer RQ3 (\textit{``What are the main challenges and open research problems in the development of next generation, robust software platforms for Smart Cities?''}), we analyzed the challenges pointed out by Smart City research papers. Table \ref{tab:three} presents an overview of the main challenges, which we describe in the following.

\begin{table}[!htbp]
\tbl{Overwiew of most cited Challenges and Open Research Problems\label{tab:three}}{%
\label{tab:challenges}
\begin{tabular}{|p{3cm}|p{6cm}|p{4cm}|}
\hline
\textbf{Challenge} 
& \textbf{Description} 
& \textbf{Technologies/Tools}
 \\\hline

Privacy & Protecting data collected from citizens, city, and enterprises. & Cryptography, Anonymization, and Access Tokens \\\hline  

Data Management & Managing all the data collected in the platform & NoSQL and Relational Databases and processing tools.  \\\hline

Heterogeneity & Ensuring the interoperability of devices \mbox{and applications} & Standards, Ontology, and a City Unified Model. \\\hline

Energy Management & Managing the electricity used by devices deployed in the city.  & \\\hline

Communication & Enabling communication among heterogeneous devices. & M2M techniques. \\\hline

Scalability & Allowing the growth of devices and users connected to the platform. & Distributed tools and algorithms and P2P applications. \\\hline

Security & Protecting the city data, services, and infrastructure. & Cryptography, Access Tokens and Devices.\\\hline

Lack of Testbeds & There are not sufficient testbeds to experiment Smart City solutions. & Simulators. \\\hline

City Models & Defining a model describing the city. & Semantic Web and Ontologies. \\\hline

Platform Maintenance & Maintaining the city systems and infrastructure. & Monitoring and Alert tools. \\\hline

\end{tabular}}
\end{table}%

\begin{itemize}

\item \textbf{Privacy:} is the most cited challenge to implementing a Smart City platform; the main reason pointed by Hassan et al. \cite{HealthcareHassanChallenges} and Balakrishna \cite{balakrishna2012enabling} is that the data collected from the city includes personal, enterprise, and governmental data that should not be accessed by other unauthorized users. Wan et al. \cite{wan2012m2m} discuss legal problems in using data belonging to platform users. 

\item \textbf{Data Management:} Many authors also cite data management as a challenge, because the platform has to store and process a large amount of data and use efficient and scalable data storage and processing algorithms \cite{su2011scandapplications,smartCityTransportation,perera2014iot}. Data Analysis is also a challenge, because it is hard to extract useful knowledge \cite{HealthcareHassanChallenges}. Another challenge is data trustworthiness; for example, Wu et al. \cite{chao2014concinnity} claim that  a large number of data sources make it difficult to ensure that all the data are correct. 

\item \textbf{Heterogeneity:} This is a challenge because of the differences between the devices in a Smart City, and the difficulty of relating data from different sources \cite{chao2014concinnity,su2011scandapplications,wan2012m2m}. Naphade et al. \cite{naphade2011innovation} raise the problem of managing data across all city systems because of variations in data from different sources. Other authors \cite{wenge2014scarchitecture} state that a Smart City platform has to define standards across heterogeneous devices, systems, and domains. 

\item \textbf{Energy Management:} Some authors cite Energy Consumption as a challenge to be faced by all the components of the platform, such as sensors, actuators, and servers \cite{perera2014iot}. Moreover, Hassan et al. \cite{HealthcareHassanChallenges} point out that energy management in a Smart City health care application is also important, because applications or services in domains like this cannot fail due to power outages. 

\item \textbf{Communication:} Since the smart cities of the future will incorporate a massive amount of devices, enabling communication among these devices will be a challenge. Some authors \cite{wan2012m2m,HealthcareHassanChallenges} discuss the domains in a Smart City that depend on mission-critical communication to ensure reliability, such as health care and public safety. In addition, Djahel et al. \cite{smartCityTransportation} explain that good communication mechanisms are required to share platform data with applications. 

\item \textbf{Scalability:} Within the next decades, the number of connected devices in a Smart City will continually increase \cite{balakrishna2012enabling}, requiring a strong level of scalability in the associated software platform. Moreover, the number of users, services, and data stored will increase with population growth and on special events in the city. Su et al. \cite{su2011scandapplications} discuss how a Smart City platform must support large-scale, efficient services. As an example, Sinaeepourfard et al. \cite{estimatingSensorsBarcelona} estimated that the city of Barcelona will need more than 1 million sensors to cover all the city, generating more than 8 GB of data every day.

\item \textbf{Security:} Unauthorized users accessing city services without permission may cause a lot of harm. Hancke et al. \cite{hancke2012role} consider whether city networks will be safe from cyber-terrorism and cyber-vandalism. Gurgen et al. \cite{gurgen2013cpssc} highlighted the importance of security in CPS platforms, as such systems control part of the city infrastructure, which a malicious user can corrupt, e.g., by tampering with traffic lights and light posts. 

\item \textbf{Lack of Testbed:} The lack of testbeds is cited by Elmangoush et al. \cite{elmangoush2013openmtc} and Muñoz et al.\cite{munoz2011forefront} as a challenge to the development of platforms for Smart Cities. Without testbeds, it is hard to perform tests and experimentation to discover the real challenges that deploying a Smart City platform will present. Smart City Simulators \cite{santana2016simulator} could be a much lower-cost alternative for experimentation.

\item \textbf{City Models:} Some authors also argue that it is hard to understand a city and describe an effective and efficient model for it. For example, Wu et al. \cite{chao2014concinnity} claim that it is necessary to create a useful model of the city to make intelligent decisions. Naphade et al. \cite{naphade2011innovation} state that modeling is required to observe and understand the city activity and to avoid generating unnecessary and empty models. Mu\~noz et al. \cite{munoz2011forefront} state that a unified model of the city is required, so that the huge amount of heterogeneous data generated can be shared among applications and services.

\item \textbf{Platform Maintenance:} Three works state that deploying and maintaining the platform is a challenge. Perera et al. \cite{perera2014iot} discuss the difficulty of maintaining a middleware to manage millions or billions of devices connected to the platform. Similarly, Wenge et al. \cite{wenge2014scarchitecture} discuss that the administration of the platform can be a challenge, due to its size as well as the very large number of devices spread across the city. Hancke et al. \cite{hancke2012role} point out that addressing coordination issues in the sensor nodes can be a problem, again because of the size of the city sensor network. 

\end{itemize}

\section{Reference Architecture for Smart City Platforms}
\label{sec:reference-architecture}

Based on the knowledge surveyed in this paper, we present a novel, comprehensive reference architecture to guide the development of next-generation software platforms for Smart Cities. The platform was derived from architectures proposed in previous works, with enhancements based on the requirements and challenges described in this survey. First, we describe and analyze the architecture of two platforms presented in the literature: CiDAP and OpenIoT. Then, based on these early works, and on the answers to the research questions presented before, we derived a novel reference architecture. Finally, we compare our proposal with the other two architectures.

\subsection{CiDAP}

\begin{figure}[!htb]
\centering
\includegraphics[scale=0.5]{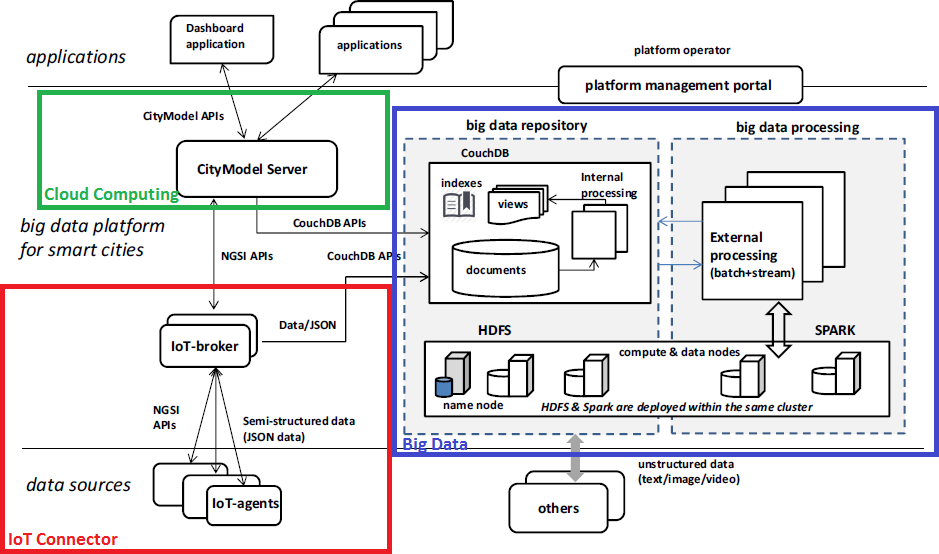}
\caption{CiDAP Platform \cite{cheng2015bigdataplat}}
\label{figura:cidap}
\end{figure}

The City Data and Analytics Platform (CiDAP) is a Big Data based platform that aims to use the data collected from the city to enable context-awareness and intelligence in applications and services. This platform processes large datasets collected from an IoT Middleware. Figure \ref{figura:cidap} presents the architecture of the platform, which has the following five main components.

\begin{itemize}

\item \textbf{IoT-Agents} connect to the IoT middleware and serve as a gateway to the devices available to the platform. Each data source of the IoT middleware is mapped to an IoT-Agent.

\item \textbf{IoT-Brokers} act as a unified interface to the IoT agents, facilitating access to the middleware data. This component communicates with the Big Data Repository to send data to be stored, and with the CityModel Server to send data to be used directly by applications.

\item The \textbf{Big Data Repository} stores raw data collected from the city and processed data from the Big Data processing component. The platform uses the CouchDB\footnote{http://couchdb.apache.org} NoSQL database, which stores data as JSON documents. This component also has an internal processing tool that makes processing simple, such as transforming data into new formats, or creating new structured views and tables to index data. 

\item \textbf{Big Data Processing} is responsible for complex or intensive processing using the data stored in the Big Data Repository, such as data aggregation or data mining. Also, it processes historical data using batch processes, or real-time data using data streams. This component uses Apache Spark for this processing. 

\item \textbf{City Model Server}  is the interface of the platform to external applications. The CityModel API allows applications to perform simple queries, complex queries, and subscribe to specific pieces of data from the platform. Simple queries request the latest data from devices, complex queries request aggregated historical data, and subscription is a mechanism for applications to periodically receive data from the devices. 

\end{itemize}

The red, green, and blue boxes in Figure \ref{figura:cidap} highlight the concepts used to implement each layer of the platform. The IoT Connector box has components to facilitate access for IoT devices in the platform. The Big Data box has components to store and analyze the data gathered from multiple sources. Finally, the Cloud Computing box indicates the interface of the platform with external applications, which is implemented using cloud services.

CiDAP is mainly concerned with storing and processing a large amount of data in the platform, which is important because of the massive amount of data collected in a city. The strong points of its architecture are data storage and processing, the real-time and batch processing modules, and the fact that the associated platform was already tested in the SmartSantander testbed.

An important limitation of CiDAP is that the platform does not foresee specific services and tools for application developers, and does not allow the deployment of new services in the platform, making its extensibility difficult.

The red, green, and blue boxes in the figure highlight the concepts used to implement each layer of the platform. The IoT Connector box has components to facilitate the access of IoT devices in the platform. The Big Data box has components to store and analyze the data gathered from multiple sources. Finally, the Cloud Computing box indicates the interface of the platform with external applications which is implemented using cloud services.

CiDAP is mainly concerned with storing and processing a large amount of data in the platform. It is important because of the huge amount of data collected in a city. The strong points of its architecture are the data storage and processing, the real-time and batch processing modules, and the fact that the associated platform was already tested in the SmartSantander testbed.

An important limitation of CiDAP is that the platform does not foresee specific services and tools for application developers and does not allow the deployment of new services.

\subsection{OpenIoT}

\begin{figure}[!htb]
\centering
\includegraphics[scale=0.4]{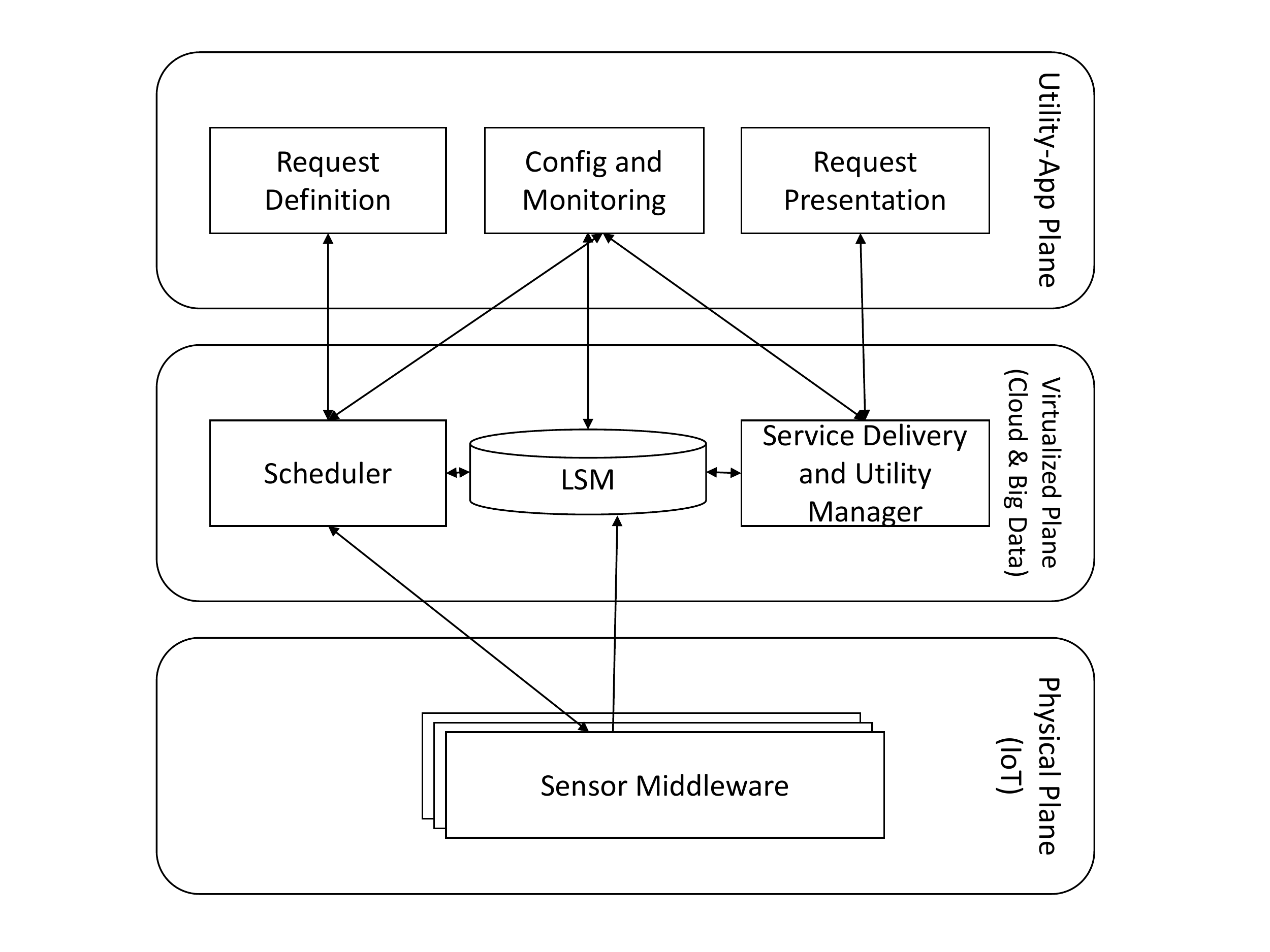}
\caption{OpenIoT Platform \cite{petrolo2014cloudofthings}}
\label{figura:openiot}
\end{figure}

OpenIoT is an Internet of Things platform used by the Vital project \cite{petrolo2014cloudofthings}  to create a Smart City platform. Figure \ref{figura:openiot} presents an overview of the platform architecture, which has three layers: the Physical Plane, the Virtualized Plane, and the Utility-App Plane. 

The Physical plane is a middleware responsible for collecting, filtering, combining, and cleaning data from sensors, actuators, and devices. This plane acts as an interface between the physical world and the OpenIoT platform. The current version of OpenIoT uses the X-GSN middleware \cite{calbimonte2014xgsn}, an open-source middleware for managing, monitoring, and controlling IoT devices.

The Virtualized plane aims to store data, execute services, and schedule the execution of these services. The main components of the Virtualized plane are the following:

\begin{itemize}

\item The \textbf{Scheduler} receives requests for services and ensures the access to resources that the service needs, such as data and data streams. This component is responsible for discovering the sensors required for a service execution. 

\item The \textbf{Cloud Data Storage} keeps all the data from the platform, e.g., data streams collected from the sensors and the data created within the platform, such as user profiles, service definitions, and registered applications. For storing data collected from the IoT middleware, OpenIoT uses the Linked Sensor Middleware (LSM) \cite{le2012middleware}.

\item The \textbf{Service Delivery and Utility Manager} has three primary functions: handling the combination of the data collected from the IoT middleware, allowing service definitions, and delivering the results of requested services to the platform or to third-party applications. Also, this component keeps track of the usage of the services defined in the platform for accounting and billing.

\end{itemize}

The Utility-App Plane, the user interface of the platform, has three main components:

\begin{itemize}

\item \textbf{Request Definition} enables users to define new applications using the services deployed on the platform, including the definition of service mash-ups.

\item \textbf{Request Presentation} executes the applications created in the Request Definition component. When a user executes an application, it communicates with the Service Delivery and Utility Manager to retrieve the results from the service executions.

\item \textbf{Configuration and Monitoring} allows configuration of platform parameters, such as periodicity of sensor data reads and monitoring the health of all platform devices and components.

\end{itemize}

OpenIoT is a complete platform, handling almost all the main requirements that we described in the survey. The strong points of this platform are the use of an IoT middleware to configure and collect data from devices, the middleware to store the data collected from sensors, the development tools, and the fact that the platform is open source. However, its architecture does not consider other data sources, such as social networks, and does not provide support for pre-processing services relevant when dealing with Big Data.

\subsection{The Unified Reference Architecture}

\begin{figure}[!htb]
\centering
\includegraphics[scale=0.5]{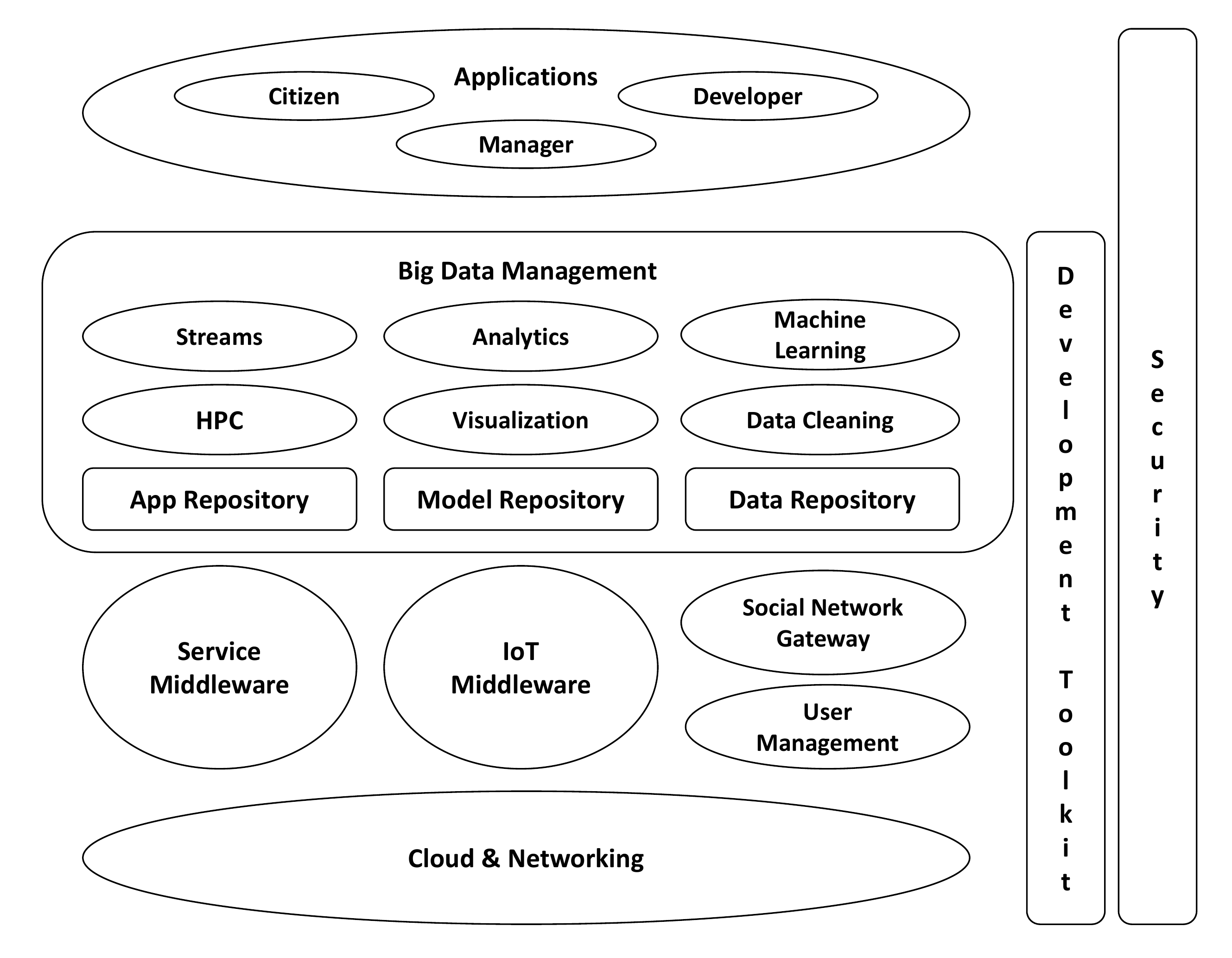}
\caption{Reference Architecture for Smart City Platforms}
\label{figura:reference_architecture}
\end{figure}

Based on the answers to the research questions of this survey, the 23 platforms analyzed, and on the two architectures presented above, we derived a novel reference architecture for Software Platforms for Smart Cities. With this reference architecture, we answer the general research question stated in Section \ref{sec:introduction} (\textit{``What are the elements required for the development of a highly-effective software platform for enabling the easy construction of highly-scalable, integrated Smart City applications?''}). Figure \ref{figura:reference_architecture} presents an overview of the architecture.

The lowest level component of the reference architecture is \textbf{Cloud and Networking}, which is responsible for the management and communication of the city network nodes. This component has to identify all the devices connected to the platform, including servers, sensors, actuators, and user devices. Using cloud computing concepts is important to ensure some fundamental non-functional requirements, including scalability and extensibility.

Just on top of the Cloud and Networking infrastructure, the reference architecture includes the \textbf{IoT Middleware} and the \textbf{Service Middleware}. The former has to manage the city IoT network and enable the effective communication of the platform with the user devices, city sensors, and actuators. The Service Middleware has to manage the services that the platform will provide to the applications, performing operations such as publishing, enacting, monitoring, composing, and choreographing these services.

The X-GSN middleware can be used to implement the IoT Middleware, which is already used in the OpenIoT project. Another option is to use components of the Sentilo platform, which is also open-source, and implement a complete IoT middleware. The CHOREOS framework \cite{issarny2011jisa} can be used to implement the Service Middleware; this project aims to enable choreographies of large-scale service-based software systems.

To provide better services to the citizens, it is important for the platform to store some user data and preferences, which is the role of the \textbf{User Management} component. But, to ensure user privacy, this data must be properly protected, and permission to store it must be acquired from the user. Moreover, as the city platform will have many applications, it can be helpful to offer a single sign-on mechanism.

Social networks will have a major role in Smart Cities. They can be used to retrieve data from city conditions, and can be an efficient communication channel between the platform and city government with the citizens. Therefore, it is important to allow the integration of the Smart City platform with existing social networks. This is the role of the \textbf{Social Network Gateway}. To implement this gateway, many tools can be used, such as Spark Streaming, which reads data streams of Twitter, and Spring Social, which is a Java-based framework to facilitate the connection with social networks such as Twitter, Facebook, and LinkedIn.

\textbf{Big Data Management} is a module to manage all the data in the platform. It is responsible for storing the data collected from the city and generated by the platform. To this extent, the reference platform has three repositories: (1) an \textbf{App Repository} to store applications, including its source/binary code, images, and associated documents; (2) a \textbf{Model Repository} to store the city models, such as a traffic model, sensor network model, data model, city maps, and an energy distribution model; and (3) a \textbf{Data Repository} to store the data collected from sensors, citizens, and applications. Because of the amount of data that a platform for Smart Cities generate, NoSQL databases can be more suitable than relational databases.

Besides the data storage, the Big Data Management module is also responsible for the processing of city data. There are two types of data processing that might be more suitable for different situations: \textbf{Stream processing}, to perform real-time analytics and data-flow processing; and \textbf{Batch processing}, to analyze large data-sets. Moreover, this module must be capable of performing useful pre-processing tasks, such as data filtering, normalization, and transformation.

The Big Data module also has a \textbf{Machine Learning} component, which facilitates understanding of the city by automatically building models of city processes behavior and making predictions of city phenomena. Since a Smart City will produce an enormous amount of data, a \textbf{Data Cleaning} component is responsible for garbage collection, deleting unneeded data, and archiving old data on slower, high capacity data stores.

To implement the Big Data Management components, many open-source tools are available. To the repositories, NoSQL Databases, such as CouchDB, MongoDB, and Cassandra, can store the unstructured or semi-structured data, such as sensor reads and social networks posts. Relational Databases, such as MySQL and PostgreSQL, can store structured data, such as user information and the platform configuration.

To implement the processing engines, many tools are also available. To execute batch processing, Apache Hadoop and Apache Spark are widely used by other platforms. Apache Spark also provides a stream data processing tool, likewise Apache Storm. Many tools offer machine learning algorithms to process large data sets such as Weka\footnote{Weka - https://weka.wikispaces.com/}, Spark MLib, and Scikit-Learn\footnote{Scikit-Learn - http://scikit-learn.org/stable/}.

Relying on aforementioned middleware component, application developers and smart city operators will develop and deploy Smart City applications. By using open data and open services provided by a city, common citizens and users may also execute, or even develop, novel applications to run on top of the city's smart infrastructure. The applications will use the services and data from the platform, but also generate and store data on the platform. The platform should provide an SDK to facilitate the development of applications, including tools such as an Integrated Development Environment (IDE), libraries, and frameworks for commonly used programming languages, and a Smart City Simulator for debugging and experimenting with applications before real deployment.

All components of the platform must support several non-functional requirements, such as scalability, security, privacy, and interoperability. Scalability is fundamental because of the huge amount of devices, data, and services in the platform. Privacy and Security are important because the platform collects, stores, and processes sensible data from the city and citizens. Interoperability will allow the integrated operation of different types of services, devices, and applications. Table \ref{tab:tech_implement} presents options to implement the reference architecture using tools that the platforms described in the survey use.

\begin{table}[!htbp]
\label{table:technologies_other_platforms}
\tbl{Technologies used in the platforms implementation\label{tab:tech_implement}}{%
\begin{tabular}{|l|l|}
\hline

Component & Tools \\\hline 
IoT Middleware & Sentilo and X-GSN\\\hline 
Data Repository & MongoDB, CouchDB, MySQL, IBM DB2, and Redis \\\hline 
Data Processing & Spark and Hadoop \\\hline 
Stream Processing & Storm \\\hline 
Cluster Management & Apache ZooKeeper and Haddop YARN\\\hline 
Cloud Environment & OpenNebula and Microsoft Azure \\\hline 
Data Access & REST APIs and Jersey \\\hline 
Security & SAML Protocol  \\\hline 
Machine Learning & Weka, Spark MLib, and Scikit Learn \\\hline 

\end{tabular}}
\end{table}%

\subsection{Comparison of Architectures}

In our architecture, we combined aspects of both platforms described in the beginning of this section. Our Big Data module is similar to the one in CiDAP; both foresee batch and real-time processing and big data storage components. However, we added the idea of an application repository (to store data and meta-data associated with applications so that we can better manage and reflect on the applications executing in the city), as well as a model repository (to store different types of models associated with various city-related phenomena such as different kinds of maps, data flows, user behaviors, automated processes, and more).

Similar to OpenIoT, we included a Cloud and Networking layer to manage the devices that collect data from the city and execute service and application components. We also included a service middleware to support many service-related operations, such as deployment, management, composition (via orchestrations and choreographies), and enactment; OpenIoT also provides a Service Delivery component with more limited support for some of these operations.

We also included some components that are not in these two architectures, but were considered relevant in our literature review. The first is the Social Network Gateway, which is important because social networks connect citizens, the city administration, and service providers, and generate a lot of useful data for city applications. 

Although OpenIoT provides some development tools to create applications using the available services, a Smart City platform will have to provide a complete software development toolkit. This SDK has to be aware of all the components of the platform, and enable the construction of sophisticated mash-ups based on them. For example, it must allow the development of a service using data from the IoT middleware, combined with data from social networks, and thereby generating a data stream that is filtered, processed, distributed to other users that have subscribed to a specific channel and, later, summarized and stored in a long-term persistent storage in order to maintain historical records.

\section{Discussion}
\label{sec:discussion}

We now discuss the findings of this research. Section \ref{subsec:tec-req} relates the four enabling technologies with the functional and non-functional requirements; Section \ref{subsec:challenges-discussion} discusses open research challenges; Section \ref{subsec:implications} presents the implications of our survey to Smart City stakeholders, such as city managers, citizens, and developers. Finally, Section \ref{subsec:limitations} considers the limitations of this work.

\subsection{Enabling Technologies and Requirements}
\label{subsec:tec-req}

This survey presented multiple approaches for the development of Smart City platforms. From this study, four highly significant functional requirements emerged: management of sensor and actuator networks; management of the data collected from the city; provisioning, management, and development of services; and an environment for the development and deployment of Smart City applications. These features can be related to the enabling technologies, mapping them onto the major functional and non-functional requirements of Smart City platforms.

Technologies around the Internet of Things are used for managing the sensor and actuator networks and their challenges, such as heterogeneity, scalability, and adaptation. Big Data and Cloud Computing are used to deal with the massive amount of data generated from multiple data sources in the city, such as WSN, social networks, and user devices. Big Data tools are required for most data-related activities, such as storing, analyzing, and sharing. Cloud Computing provides a scalable and elastic environment to store and process city data.

\begin{figure}[!htb]
\centering
\includegraphics[scale=0.4]{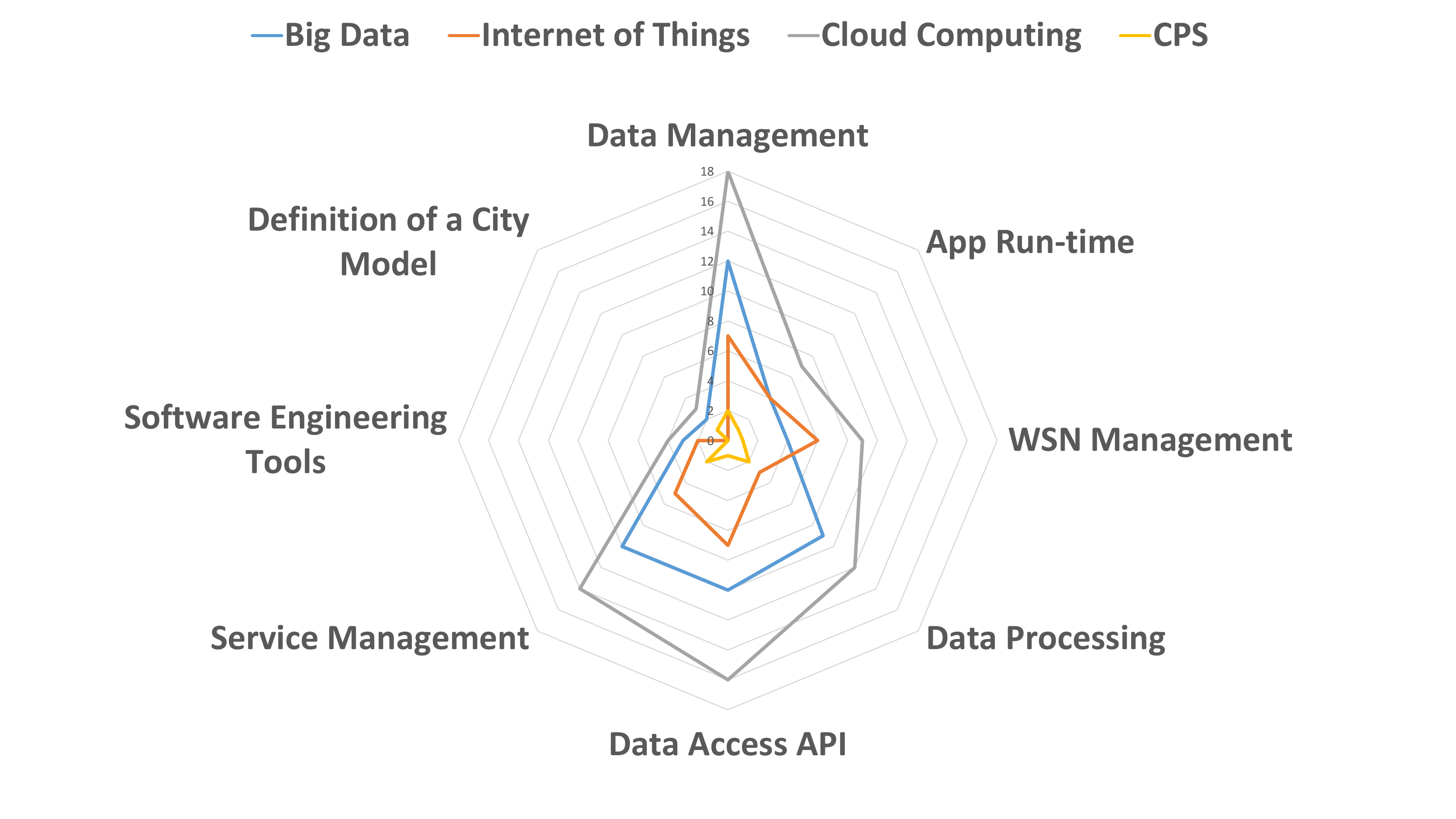}
\caption{Relationship between functional requirements and enabling technologies}
\label{figura:req-tech}
\end{figure}

Figure \ref{figura:req-tech} shows the relation between the implemented functional requirements from platforms and the enabling technologies. For example, it is possible to verify that most of the Big Data platforms handle Data Management and Data Processing. Cloud Computing platforms handle External Data Access and Service Management.
 
\begin{figure}[!htb]
\centering
\includegraphics[scale=0.4]{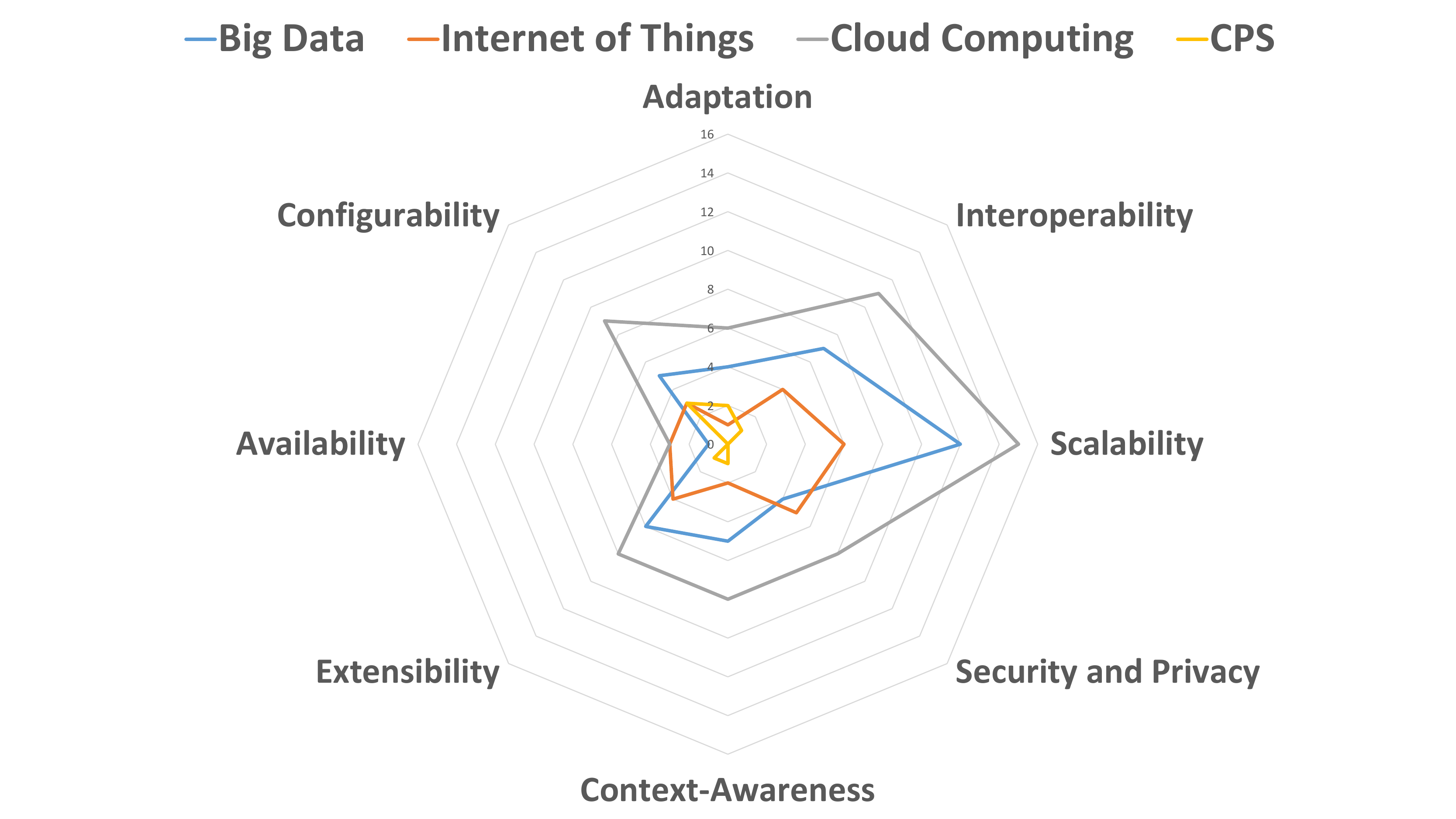}
\caption{Relationship between non-functional requirements and enabling technologies}
\label{figura:non-func-req-tech}
\end{figure}

Figure \ref{figura:non-func-req-tech} relates the non-functional requirements and the enabling technologies. We can observe that most platforms are concerned with scalability, regardless of the enabling technology used. It is possible to verify relationships between other non-functional requirements and the technologies. For example, all the CPS platforms handle configurability. Extensibility is mostly offered by platforms that use Big Data, and interoperability mostly by platforms that use IoT.

\subsection{Challenges and Open Research Problems}
\label{subsec:challenges-discussion}

Most of the significant challenges and research problems in implementing a platform for Smart Cities is related to data management. The most cited problem in the literature is ensuring the privacy of user data, because of the amount of personal and critical data that a platform needs to handle, such as user locations and medical records.

The second most cited challenge is heterogeneity, because of the large number of different systems, services, applications, and devices that a platform must support. We were surprised that only three authors cited scalability as a problem, because it certainly will be a great challenge support the huge number of devices, users, data, and services in a large metropolis.

An important and understudied issue is how to create a generic platform to support the requirements of different cities. Some literature focuses on a particular city, such as WindyGrid, SmartSantander, and Padova Smart City. Other platforms provide solutions without discussing the characteristics of the cities in which that solution should be applied. The studies proposing generic solutions for Smart Cities lack a discussion concerning how the components of these platforms could be adapted to fit cities of different sizes and characteristics.

\subsection{Implications}
\label{subsec:implications}

This paper presented important features that software platforms for Smart Cities should handle. The results give important references for several city stakeholders, such as platform developers, application developers, city managers, system operators, end-users, and Smart City researchers. In this section, we discuss the potential implications of our findings for these stakeholders.

The enabling technologies highlight the infrastructure needed to build Smart Cities. City managers can use this information to improve their investment decisions. Big Data and Cloud Computing deal with an enormous volume of data storage and network infrastructure to access data and services. The city must be equipped with sensors, actuators, and Internet services to take advantage of the Internet of Things and Cyber-Physical Spaces. Besides, the survey can help Smart City application and system developers decide what technologies to use.

The reference architecture highlights the functional and non-functional requirements that platforms and applications developers should consider when developing software for Smart Cities. For platforms developers, this survey indicates that is necessary to deal with big heterogeneous and distributed systems, as well as critical and personal data, in an effective and efficient way. For application developers, the reference architecture shows what kind of services and data they can use to provide better experiences to their end-users. By discussing examples of these systems, we show to end-users, or citizens, the range of system domains that can be developed to facilitate their daily routine, such as urban mobility, air pollution, and heath care.

Finally, this survey can also helps Smart City researchers by discussing the main open research questions and challenges to be overcome to build smarter cities. These challenges can guide future work in this research area.

\subsection{Limitations}
\label{subsec:limitations}

In this survey, we decided to describe only the most cited enabling technologies used by Smart City platforms. However, we found other less employed technologies, such as M2M Communications and the Semantic Web. These non-cited technologies are used by few platforms, or are used to solve a small problem but not to serve as a fundamental architectural component of the platform. Thus, there might be key technologies that end up being very relevant in the future that have not yet been identified in this survey.

We used the most cited paper of each research project to extract components, requirements, and features of the platforms. Other papers, or the project website, may define different aspects.

In this research, we classified the papers according to the enabling technologies only when they were explicitly mentioned. However, we noticed that, in some papers, they were pointed out as a motivating aspect or future work. For example, Khan et al. \cite{khan2013bigdataanalytics} do not explicitly mention IoT in the architecture, but the authors discuss the possibility of using smart hardware such as sensor networks or smart household appliances, which can be organized in an IoT system.

\section{Related Surveys}
\label{sec:related-works}

In our literature search, we found four papers that also surveyed platforms and applications for Smart Cities.

Da Silva et al. \cite{da2013smart} surveyed architectures of Smart Cities platforms, analyzing the requirements handled by the platforms. However, they analyzed few platforms and did not distinguish functional and non-functional requirements, and did not address future research and open challenges in the area.

Yin et al. \cite{yin2015surveySmartCities} conducted a survey on Smart Cities. Although the paper presents some platforms, the main goal of their work was to understand the concept of Smart Cities, identifying the enabling technologies and Smart City research issues.

Al Nuaimi et al. \cite{applicationsBigData} reviewed the use of Big Data tools and concepts in applications for Smart Cities. The paper mainly presents the relation between the challenges to creating applications for Smart Cities and the use of Big Data tools. It also identifies Smart City requirements that Big Data tools can address. It has some similarities with our work, but we conducted a more general and comprehensive survey.

Finally, Botta et al. \cite{botta2015integration} presented a study of the integration of Cloud Computing and the Internet of Things, defining this novel paradigm as CloudIoT. They describe applications that use this paradigm, such as health care, transportation, and smart cities. The paper presents platforms that use the two concepts, some of which are also presented here, such as OpenIoT and ClouT.

In our work, we studied Smart City software platforms and the related ICT problems, aiming to derive the major functional and non-functional requirements, and the technical and research open challenges. Besides, we presented a reference architecture derived from the requirements pointed out by the surveyed studies.

\section{Conclusion}
\label{sec:conclusions}

Smart City is a concept that has gained increased attention in academic, industrial, and governmental circles. While the urban population is growing, the infrastructure and resources required to support citizens are often insufficient, leading to a degradation in public services. Information and Communication Technologies provide important tools to reduce this problem, helping to improve the sustainable use of resources, city services, and the citizens' quality of life.

Using a software platform rather than \textit{ad hoc} solutions is a more robust and sustainable way to support the features needed by a Smart City environment. In this paper, we surveyed the current research on Smart Cities platforms, aiming to discover theirs most relevant requirements and how to facilitate the development, integration, and deployment of Smart City applications. We analyzed 23 studies from different groups, proposing multiple approaches for the development of a software platform to answer our general research question ``What are the elements required for the development of a highly-effective software platform for enabling the easy construction of highly-scalable, integrated Smart City applications?''

Based on the analyzed projects, we derived a unified reference architecture supporting the main requirements needed to build a software platform for Smart Cities. Thus, this paper contributes to the state-of-the-art by providing a guide to help software developers and city managers determine the necessary components to handle the functional and non-functional requirements of a software platform for Smart Cities.

The reference architecture is based on the answers of the three research sub-questions. RQ1 (``What are the enabling technologies used in state-of-the-art software platforms for Smart Cities?'') showed us that the Internet of Things, Cloud Computing, Big Data, and Cyber-Physical Systems are the most cited enabling technologies. Answering RQ2 (``What are the requirements that a software platform for Smart Cities should meet?''), we could relate these technologies to the requirements that a software platform should handle. For example, most of the Big Data platforms mention Data Management as a requirement, while Configurability is strongly related to CPS platforms. In this way, an important contribution of this survey is to discuss the requirements that need to be implemented when using a specific enabling technology. In contrast, it helps to decide which technology to use when a specific functional or non-functional requirement is desirable.

Finally, to answer RQ3 (``What are the main challenges and open research problems in the development of next generation, robust software platforms for Smart Cities?''), we presented the most cited challenges and open research problems, according to the literature. These challenges were considered when deriving the reference architecture. In this sense, an important contribution of this survey, especially for developers and researchers of software platforms, is to identify which platform components should be the focus of future work.

This survey described several Smart City initiatives, but all of them are still in their initial phases, posing multiple challenges and open problems that need to be addressed. A collaborative effort of research groups, commercial companies, NGOs, and governments is required to tackle the multitude of scientific, technical, political, and social problems related to the establishment of really-smart cities, reaching the ultimate goal of improving the quality of life of all of a city's citizens, irrespective of its social and financial situation.

\begin{acks}
This publication is the result of a project promoted by the Brazilian Informatics Law (Law No. 8248 of 1991 and subsequent updates) and was developed under the Cooperation Agreement 073/2016 between Universidade de S\~ao Paulo, FDTE, and Hewlett Packard Enterprise Brazil.
\end{acks}

\bibliographystyle{ACM-Reference-Format-Journals}
\bibliography{SmartCityPlatformsSurvey.bib}

\medskip

\end{document}